\documentstyle[epsfig,graphicx]{mn}

\def\etal{{et~al.}}
\def\eg{{e.g.}}

\def\cf{{cf.}}
\mathcode`\|="8000		
{\catcode`\|=\active \gdef|#1{_{\rm #1}}}
\def\unit#1{\,{\rm {#1}}}
\def\umult#1#2{\ifx#2\unit\def\tmpa{\umulti#1#2}\else\toks0=\expandafter{#2}%
\edef\tmpa{\noexpand\umulti\noexpand#1\the\toks0}\fi\tmpa} 
\def\umulti#1#2#3{\ifx#2\unit\unit{#1#3}\else%
\message{\noexpand\umult error: must precede unit or \noexpand\unit}%
\unit{#1}#2#3\fi} 
\def\Kelv{\unit{K}}
\def\secnd{\unit{s}}
\def\yr{\unit{yr}}
\def\cm{\unit{cm}}

\def\parsec{\unit{pc}}
\def\kms{\unit{km\,s^{-1}}}
\def\Msun{\unit{M_\odot}}

\newcommand{\mn}{MNRAS}
\newcommand{\ana}{A\&A}
\newcommand{\apj}{ApJ}

\def\rd{{\rm d}}                                      
\def\rdif#1#2{\mathchoice{\rd{#1}\over\rd{#2}}{\rd{#1}/\rd{#2}}
{\rd{#1}/\rd{#2}}{\rd{#1}/\rd{#2}}}                            

\def\ion#1#2{{\rm #1}%
\ifmmode{\mathchoice{\scriptstyle}{\scriptstyle}
{\scriptscriptstyle}{\scriptscriptstyle}{\rm\uppercase{#2}}}%
\else$\,\scriptstyle\rm\uppercase{#2}$\fi}

\def\HII{{\ion{H}{II}}}

\makeatletter
\def\@cite#1#2{(\if@tempswa #2 \fi #1)}
\makeatother
\def\ee{\protect\pee}
\def\pee#1{\ifmmode{\times10^{#1}}\else$\times10^{#1}$\fi}

\title[Photoionized columns]{Hydrodynamics of photoionized columns in
the Eagle Nebula, M~16}

\author[R.J.R. Williams \etal{}]{
R.J.R. Williams, D. Ward-Thompson \& A.P. Whitworth\\
Department of Physics and Astronomy, Cardiff University, PO Box 913,
Cardiff CF24 3YB}
\date{Received **INSERT**; in original form **INSERT**}
\pagerange{\pageref{firstpage}--\pageref{lastpage}}
\begin{document}
\label{firstpage}
\maketitle

\begin{abstract}
We present hydrodynamical simulations of the formation, structure and
evolution of photoionized columns, with parameters based on those
observed in the Eagle Nebula.  On the basis of these simulations we
argue that there is no unequivocal evidence that the dense neutral
clumps at heads of the columns were cores in the pre-existing
molecular cloud.  In our simulations, a variety of initial conditions
leads to the formation and maintenance of near-equilibrium columns.
Therefore, it is likely that narrow columns will often occur in
regions with large-scale inhomogeneities, but that observations of
such columns can tell us little about the processes by which they
formed.  The manner in which the columns in our simulations develop
suggests that their evolution may result in extended sequences of
radiation-induced star formation.
\end{abstract}

\section{Introduction}

\HII\ regions are common features of regions of massive star
formation, such as Orion or the Eagle Nebula, M~16, in which
ultraviolet photons from young massive stars photoionize and
photodissociate the molecular gas clouds from which the stars
originally formed.  Emission from the ionized gas, in lines such as
the Balmer series or [O{\sc\,iii}]$\lambda 5007$\AA, allows the
regions to been seen at optical wavelengths.  Images of these regions
have many different forms, although they may often be characterised as
a photoevaporating blister.

The basic stratification of gas in an \HII\ region is well understood.
Close to the central star or stars the gas is highly ionized, as a
result of the strong photoionizing continuum.  Atoms which recombine
within this region are rapidly photoionized once again, and the gas is
maintained at an equilibrium temperature close to $10^4\Kelv$.  At the
edge of this region, where the photoionizing continuum spectrum has
been essentially exhausted, is an ionization front (IF) which
separates the ionized gas from atomic gas.  The ionization front is
generally narrow, compared to the overall scale of the region.  Beyond
the IF, the gas in the photodissociation region (PDR) is kept atomic
by photons with wavelengths longer than 912~\AA\ which do not have
sufficient energy to ionize hydrogen, but are energetic enough to
dissociate molecular hydrogen \cite{bd}.  Eventually, the spectrum of
photodissociating photons is also absorbed, and the gas can remain
molecular.

However, this simple stratification belies the intricate structures
often observed in \HII\ regions.  A feature of particular interest is
the photoionized columns, or `Elephant Trunks', seen at the edges of
regions such as M~16, M~20 and NGC~3603.  These intrusions of
molecular gas into the high-pressure environment of the ionized nebula
are potentially important sites for star formation, and indeed
evidence has been mounting which associates young stellar sources with
the tips of the columns in M~16 \cite{hester96,white99}.  But the
origin, structure and evolution of these columns has been the subject
of controversy for many years.  Do they result from pre-existing dense
cores in the molecular gas or from instabilities in the IF/PDR
structure?  What is the timescale for their evolution?

\subsection{The model of White \etal}

White \etal~\shortcite{white99} have studied the structure of the
columns in M~16 using observations in a wide variety of wavebands.
They suggest that the heads of the columns are pre-existing molecular
clumps which are now in the process of dynamical collapse, driven by
an overpressure in the ionized gas around the heads, which will soon
result in the formation of stars.  In this model, gas at the heads of
the columns in M~16 started to collapse at most $10^5\yr$ ago, when the
ionization front (IF) moving away from the exciting stars reached
their surfaces.  The velocity of the shock driven into the neutral
material by the IF is approximately
\begin{equation}
v|s \simeq \sqrt{P|{sn}-P|n\over\rho|n}
\end{equation}
where $P|{sn}/k|B\sim1.2\ee8\cm^{-3}\Kelv$ and
$P|n/k|B\sim3.5\ee{7}\cm^{-3}\Kelv$ are the pressures inferred for the
shocked neutral gas and for the gas at the head of the column,
respectively, and $\rho|n \sim 2\ee{5} m|{H_2}\cm^{-3}$ is the density
of the gas in the head.  White \etal\ calculate that
$v|s\simeq1.5\kms$, and so that the shock will cross the head of the
column in $\sim1.5\ee5\yr$.

In consequence, White \etal{} argue that the columns have been subject
to the incident radiation flux for a time no longer than this.  For
instance, the parameters given by White~\etal{} imply that the flux of
gas into the shock is $\sim 5\ee{10}\cm^{-2}\secnd^{-1}$ H nuclei,
while the mass flux out through the IF is $\sim
4\ee9\cm^{-2}\secnd^{-1}$ H nuclei.  Even allowing for the different
areas of IF and shock, shocked gas should pile up between them to a
total column of $\sim10^{23}\cm^{-2}$ in $10^5\yr$. If the clump had
been subject to photoevaporation for more than $10^5\yr$, the shocked
gas should now be observable in the CO or $850\,\mu{\rm m}$ data.

White \etal{} provide secondary support for the age they infer from
detailed chemical and thermal modelling, which agrees with the
observed properties of the molecular gas, and particular the low
temperature of the densest molecular gas, and from the absence of
diagnostics of protostars within the cold clumps.  They argue that the
distance between the ionization front and shock propagating into the
clump will be far smaller than $0.2\parsec$, so the dense material
seen at the head of the column cannot itself be shocked gas.

Based on this model, they propose that the tips of the columns in M~16
may represent the best examples known to date for earliest stages of
Class 0 protostellar development, which have not yet collapsed far
enough to obey the full criteria for Class 0 \cite{andre96}.

\subsection{Other possibilities}

Despite the arguments of White~\etal{}~\shortcite{white99}, a lifetime
of $\la10^5\yr$ still seems difficult to reconcile with the dynamical
age of the other structures around the columns.  Since the density of
the ionized gas around the column is far smaller than that of the
molecular gas, the main ionization front is likely to be D-type.  As a
result, the IF and its leading shock will be moving into unperturbed
molecular material ahead at a velocity $\la10\kms$.  Indeed, since the
age of the stellar cluster is $1-2\ee6\yr$, the average velocity of
the ionization front over this lifetime can be no more than $2\kms$.
Even if the velocity of the IF were as high as $10\kms$ as it
propagated along the sides of the column, the IF must have passed the
head of the column at least $3\ee4\yr$ ago, so an age substantially
smaller than $10^5\yr$ seems unlikely.  Similarly,
Hester~\etal{}~\shortcite{hester96} argue that the distribution of
evaporating gaseous globules (EGGs) ahead of the column suggests that
they are being photoevaporated on a timescale of a few $\times10^4\yr$
as the column recedes, so the limit on their distribution relies on
this photoevaporation timescale rather than on the lifetime of the
finger itself.  Finally, the presence of strikingly similar columns in
other regions suggests that columns may survive for a significant
fraction of the total age of the \HII\ regions into which they
intrude.  None of these arguments excludes the picture developed by
White \etal: however, the upper and lower limits on the lifetime are
tight, and may come to exclude it in time.

\begin{figure}
\begin{centering}
\epsfxsize=8cm\epsffile{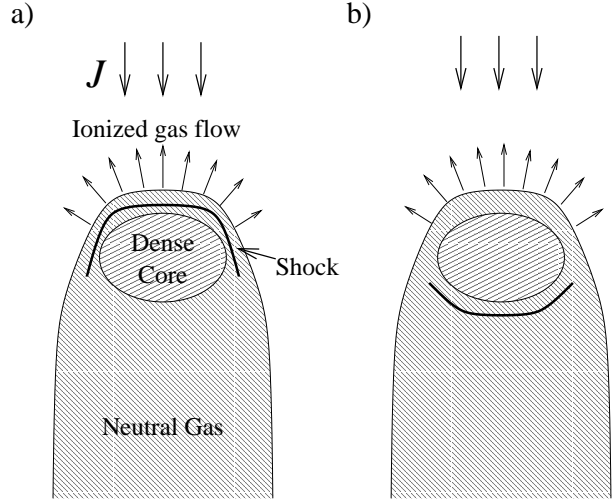}
\end{centering}
\caption{Schematic diagram of the Eagle columns, showing the possible
positions of the ionization front and shock front.  In both cases, the
ionization front is coincident with the edge of the neutral material
in the column, but in (a), the shock is between the ionization front
and the dense molecular material in the head of the column as disussed
by White~\etal~\protect\shortcite{white99}, while in (b) it is between
the head and the less dense molecular material in the barrel of the
column.}
\label{f:tips}
\end{figure}

In the present paper, we will explore longer-lived alternatives to the
model of White~\etal{} One is that the dense gas in the head is
separated from the lower density gas in the column by a shock.  The
comparison between this case and the model of White \etal{} is
illustrated schematically in Figure~\ref{f:tips}.  White \etal\ rule
out such a scenario on the basis of the narrowness of the ionization
front/shock front layer.  Indeed, the predicted flux of H nuclei
through the shock, $10^{10}\cm^{-2}\secnd^{-1}$, means that, to have
collected the observed $30\Msun$ of gas in the head, it would have
needed to propagate through approximately $1\parsec$ of gas at the
mean density of the neutral column over a timescale of $4\ee5\yr$:
substantial fractions of both the size and lifetime of M~16.  The
collapse of the neutral material in the tangential direction may,
however, decrease the time required to build up the mass
\cite[cf]{bert89}.  Structures of this form may result from density
structures with initial contrast lower than those described by White
\etal{}, or from partial shadowing of the ionization front
\cite{will99}.  The detailed structures found in numerical
hydrodynamic simulations below are more complex than those shown in
Fig.~\ref{f:tips}, but accord broadly with this scenario.

Another alternative is that the dense gas at the head of the column is
instead confined by gravity, with the shock leading the ionization
front having effectively bypassed the region and moved away
downstream.  White \etal{} find that self-gravity may be important at
the head of the column.  The size of the region dominated by the
gravity of the observed molecular core is about $\sim 0.13 (M/30\Msun)
(\sigma/1\kms)^{-2}\parsec$, where $\sigma\simeq 1\kms$ is the
effective sound speed of the gas (including contributions from
turbulence and magnetic fields as well as the thermal sound speed) and
$M\simeq 30\Msun$ is the mass of neutral gas in the head of the
column.

In the following sections, we discuss numerical hydrodynamic
simulations of these scenarios.  These have been tailored to the
central column in M~16, which has the most symmetrical appearance and
for which the observational data have the highest spatial resolution.
We follow Hester \etal's nomenclature, and term it column II (it is
referred to as $\Pi_2$ by White \etal).

The structures which we find have some similarities to the equilibrium
cometary cloud solutions discussed by Bertoldi \&
McKee~\shortcite{bm}, based on the collapse of isolated,
near-spherical clumps.  However, our time-dependent hydrodynamical
models give useful information on flow stability and the presence of
shocks.  Similar structures have been found by Garc\'{\i}a-Segura \&
Franco~\shortcite{gsf96}, in their study of the development of \HII\
regions in inhomogeneous environments, although as a result of the
global scope of their simulations individual columns were poorly
resolved.  Mellema \etal{}~\shortcite{mellea98} model the formation of
neutral tails behind isolated dense clumps in some detail.  The models
we present here have been developed for cases where the columns are
close to the edge of an \HII\ region, with parameters chosen to
compare directly with those in M~16.

In Section~\ref{s:deriv}, we discuss the derivation of suitable
parameters for our numerical models from the observational data, and
particularly the comparison of the pressures of ionized and neutral
gas across the ionization front at the head of the column.  In
particular, we find that the observational evidence for an
overpressure around the heads is weak.  In Section~\ref{s:num}, we
discuss the numerical method we have used, and, in
Section~\ref{s:cond}, the initial conditions for the simulations.  In
Section~\ref{s:res}, we present the results of these simulations, and
in Section~\ref{s:concl} discuss their implications.

\section{The Columns in the Eagle Nebula, M~16}

\label{s:deriv}

\subsection{Observational data}

\begin{figure}
\begin{centering}
\epsfysize=8cm\rotatebox{270}{\epsffile{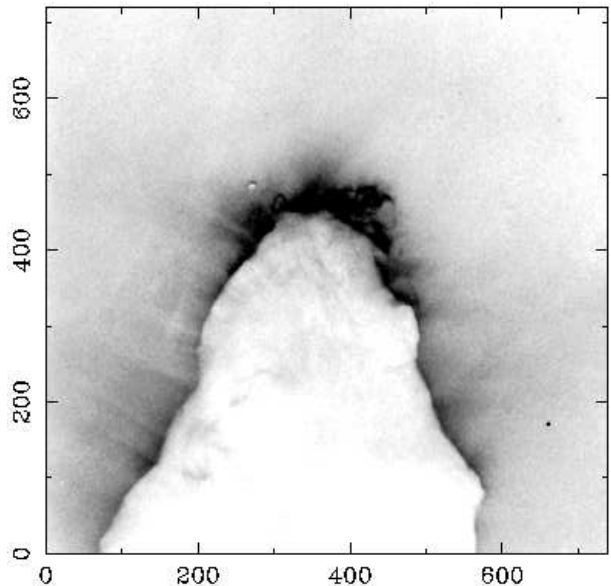}}
\end{centering}
\caption{Cleaned H$\alpha$ image, shown negative and with a grayscale
saturated in order to bring out low contrast features.  The axes are
labelled in pixels, where each pixel corresponds to $1.4\ee{15}\cm$,
for Hester \etal's adopted distance of $2{\rm\,kpc}$.}
\label{f:image}
\end{figure}

The columns in M~16 have been studied by many authors.  Perhaps the
best optical data is the HST narrow band imaging of Hester
\etal~\shortcite{hester96}.  We retrieved two H$\alpha$ images from
this data set from the HST archive, and stacked and median-limited
them to remove strong cosmic ray features and very fine-scale
structures.  The resulting image is shown in Figure~\ref{f:image}.
There are many striking irregularities in this image.  Small clumps
break the surface, and some appear fully separated from it.  Close to
the apex, the surface is puckered on a fine scale.  Some of the clumps
harbour K-band sources, and are identified as protostars by Hester
\etal{} Away from the head, the surface irregularities seem to grow in
width, with bright regions where their surfaces face towards the
source of ionization.  In addition, we note the low-magnitude
striations which are seen along lines approximately perpendicular to
the nearest surface.

White \etal~\shortcite{white99} combined this optical data with
observations of M~16 in molecular lines, millimetre and sub-millimetre
continuum and the mid-infrared, to present a remarkably complete
picture of properties of the gas and dust in and around the columns.
Within the columns, molecular gas is probed by line emission and the
associated dust is seen in millimetre/sub-mm emission, while H$\alpha$
and radio continuum observations are sensitive to structures in the
surrounding ionized gas.

Levenson~\etal{}~\shortcite{levea00} have observed excited H$_2$
emission in M~16, to study the properties of the PDR which leads the
ionization front.  They find that their observations are consistent
with a stationary photodissociation region, around $2\ee{17}\cm$ deep
(although since the properties of the incident radiation field are not
well determined, it is possible that shocks driven by the IF might
also contribute to the collisional component of this emission).

\begin{figure}
\begin{centering}
\begin{tabular}{l}
(a)\\
\epsfysize=8cm\rotatebox{270}{\epsffile{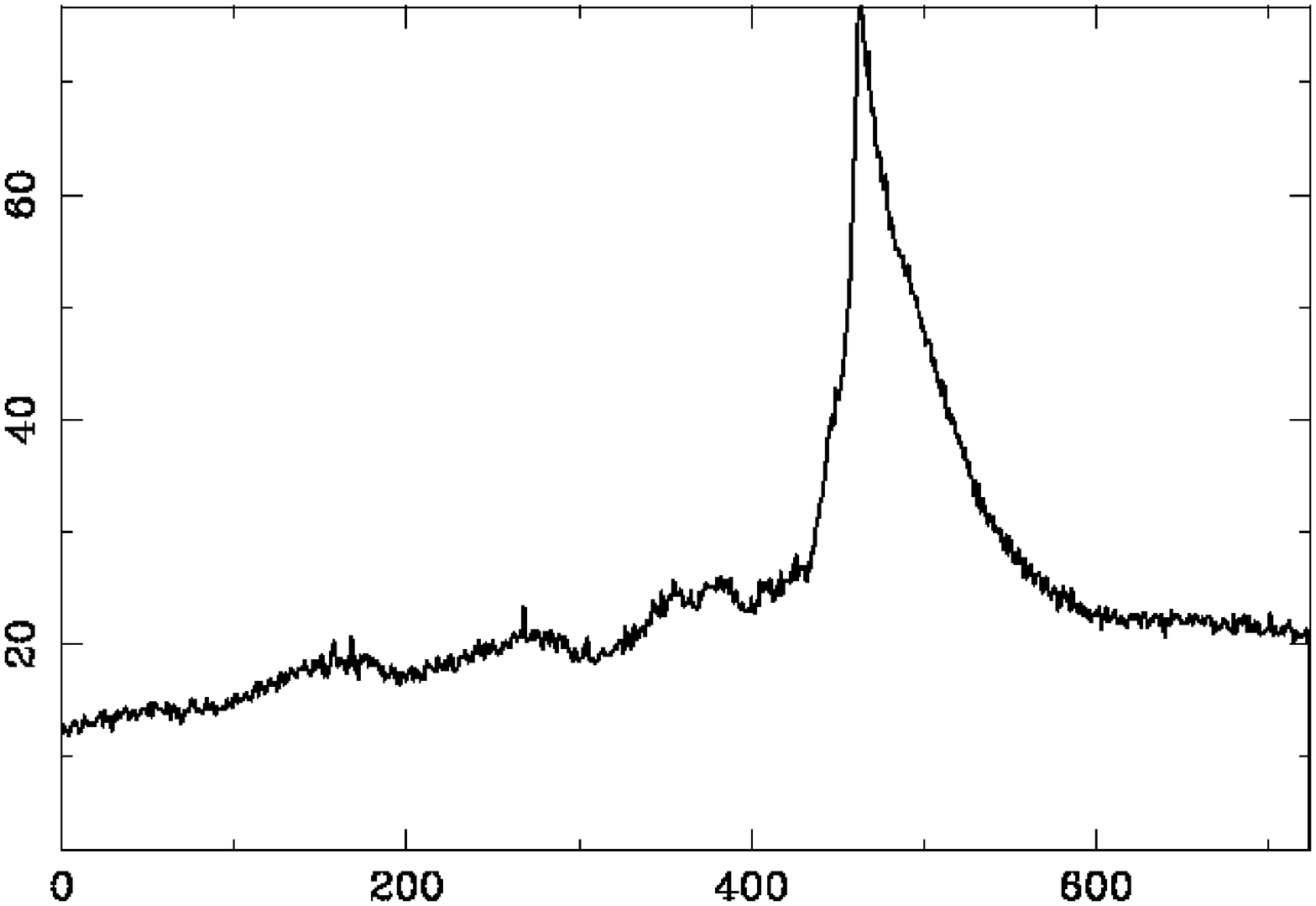}}\\
(b)\\
\epsfysize=8cm\rotatebox{270}{\epsffile{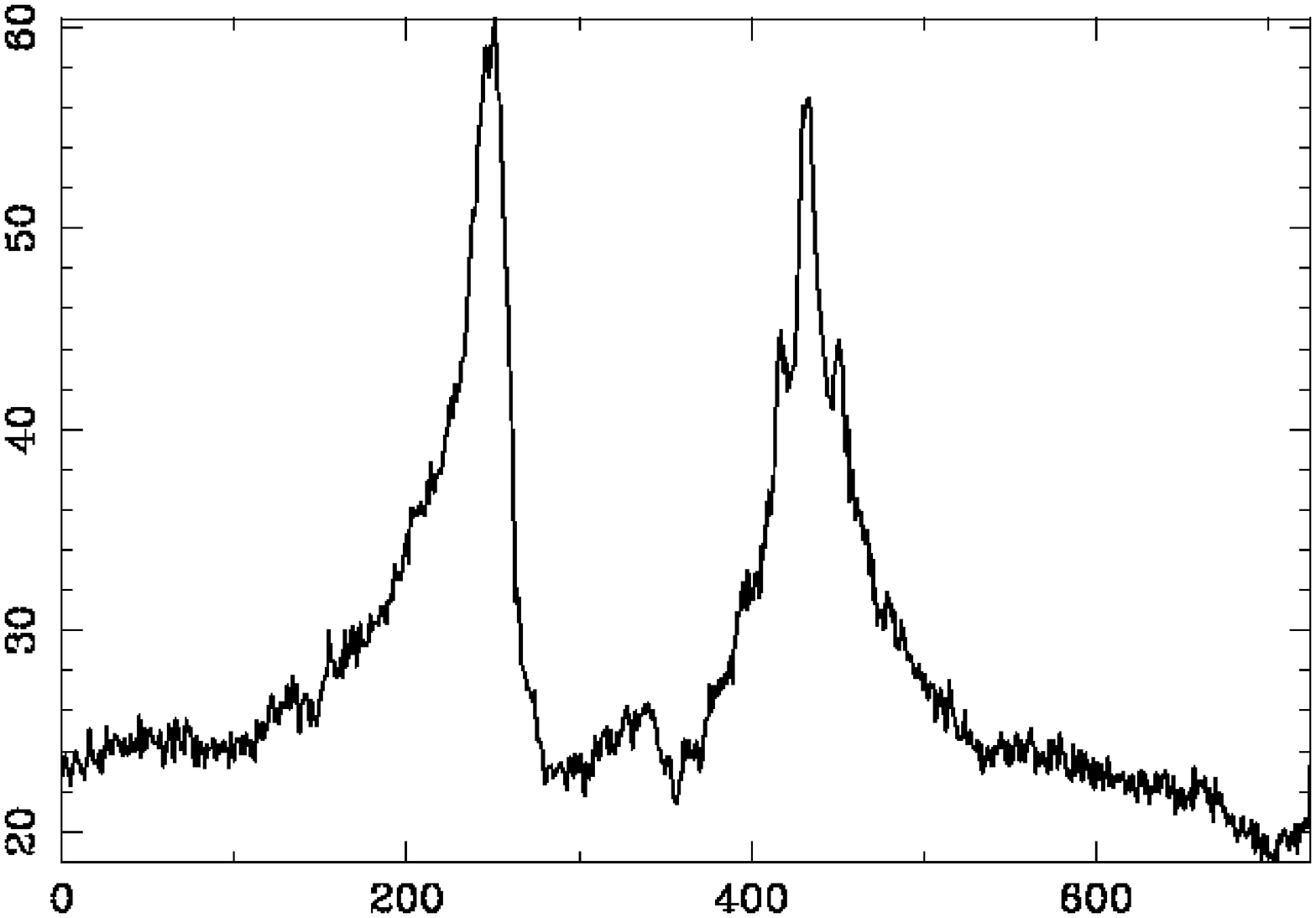}}
\end{tabular}
\end{centering}
\caption{Sections through the data (a) top-to-bottom (b) left-to-right.}
\label{f:section}
\end{figure}

\begin{figure}
\begin{centering}
\epsfysize=8cm\rotatebox{270}{\epsffile{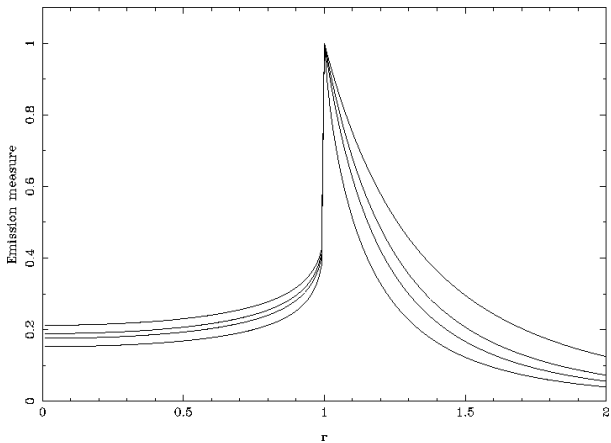}}
\end{centering}
\caption{Sections through spherical flow models for isothermal winds
which start from Mach 1, 1.2 and 2 at the surface of the column, and
for constant velocity flow.  The plots have been normalized to fixed
peak emission measure: the sharpest plot is for a Mach 1 outflow while
the smoothest is for constant velocity outflow.}
\label{f:secmodl}
\end{figure}

\subsection{Comparison with photoevaporated winds}

In Figure~\ref{f:section}, we show cuts through this image.  The first
is along the axis of the large-scale column which is vertical in
Fig.~\ref{f:image}.  The values are averages for a strip between 345
and 355 on the $x$-axis.  The second is perpendicular to this,
averaged for a strip between 395 and 405 on the $y$-axis.

In Fig.~\ref{f:secmodl}, we show the radial emission measure profile
derived from isothermal wind models, as described by Henney \&
Arthur~\shortcite{ha98}.  These authors found that such models gave a
very good fit to H$\alpha$ emission profiles of gas photoevaporated
from proplyds in the Orion nebula.  We have here been rather less
strenuous in the correction of the observational data for
contamination by continuum and [N{\sc\,ii}] emission, but our aim in
this section is simply to get an indication of the properties which
pertain at the head of column II.

Comparing the model profiles with Fig.~\ref{f:section}, we see that
their overall forms are very similar.  The central brightness of the
columns appears rather lower, relative to that at large distances,
than the model would predict, but that may be explained by the
decreasing intensity of illumination as the surface of the column
points away from the head and by the occulting of background emission
by the column.  The sharpness of the inner edge of the profile
suggests that the column is slightly closer to us than is the
photoionizing star.

More fundamentally, if (following Hester \etal) we take the radius of
curvature of the head of the column to be 90 pixels, then the breadth
of the observed H$\alpha$ profile is difficult to account for.  Not
even the most extreme of the models in Figure~\ref{f:secmodl} explains
the width of the transverse light distribution.  For comparison,
Sankrit \& Hester~\shortcite{sankh00} find that power-law models with
a radius of curvature of $2\ee{17}\cm \simeq 150$ pixels fit the
emission profile in several emission lines, which confirms that the
effective radius of curvature is larger than that apparent in the
images.  Megeath \& Wilson~\shortcite{mw97} also find that the radio
emission measure is more extended in their observations of
photoevaporating clumps in NGC~281 West than for a $n_e \propto 1/r^2$
density model.

One possible explanation for the M~16 results is that the head of the
observed column is more elongated along the line of sight than
perpendicular to it.  Alternatively, if the column is directed into
the plane of the sky \cite{pound98}, the inner peak of the emission
measure profile may be hidden from view, although the inclination does
not appear to be that extreme.  

\subsection{Inferred density and radiation field}

For the model ionized flow profiles, isothermal flow with initial Mach
numbers of $1$, $1.2$, $2$ and constant velocity flow, the effective
integration distance (i.e.\ the peak emission measure divided by the
square of the peak density) is 0.8, 1.0, 1.3, and $\pi/2$ times the
radius of curvature, respectively.  This may be compared to the value
of one-half assumed by Hester \etal{} (since the profiles are well
resolved, the correction due to resolution effects is minimal).  When
the uncertainty in the local radius of curvature is included, it seems
very likely that Hester \etal's value of $4000\cm^{-3}$ for the
electron density at the head of column II may have been overestimated.
Indeed, Hester \etal's own photoionization modelling suggested that
the density at the head of the column is $2000\cm^{-3}$ \cite[although
this has subsequently been revised by]{sankh00}.  As a result, the
ionized gas may not in fact have a greater pressure than the effective
pressure in the neutral gas at the head of the column.

The density at the head of the column can also be used to determine
the incident ionizing flux.  The thickness parallel to the radiation
field is roughly $\ell\sim0.04\parsec$ from Figure~\ref{f:section}.
Balancing the number of recombinations in the flow with the intensity
of the incident ionizing radiation, $J$, suggests that $J\sim \alpha|B
n|e^2 l \simeq 1.3\ee{11}\cm^{-2}\secnd^{-1}$, for a peak electron
density of $n|e\sim2000\cm^{-3}$.  It is difficult to follow White
\etal's analysis of their VLA data, but assuming that the integrated
$48{\rm\,mJy}$ emission corresponds to that occulted by the head of
the column, which has a radius of curvature of $0.04\parsec$
\cite{hester96}, the incident ionizing intensity predicted by the
formulae of Lefloch, Lazareff \& Castets~\shortcite{llc97} is $J\sim
1.5\ee{11}\cm^{-2}\secnd^{-1}$.  These values are mutually consistent,
but smaller than those found by White \etal, and also smaller than the
values, inferred from stellar censuses, of
$4.4\ee{11}\cm^{-2}\secnd^{-1}$ \cite{hester96} or
$2.6\ee{11}\cm^{-2}\secnd^{-1}$ \cite[from the total flux given
by]{pound98}.  However, this difference can easily be accounted for by
uncertainties in the ionizing flux of the stars, as the result of dust
absorption or photoionization between the ionizing stars and the
columns, or by the geometry of the region.

In our models, we neglect the role which a PDR may have in the
dynamics of the region.  The depth of the PDR in M~16, as derived by
Levenson \etal{}~\shortcite{levea00}, is close to the radius of
curvature of the finger tips, so this assumption is only marginally
valid.  However, since the column density inferred is close to the
stationary models \cite[rather than being substantially affected by
advection, \cf{}]{bd}, it seems likely that the jump conditions across
the combined IF/PDR structure are determined principally by the flux
incident on the IF.\footnote{If, in other cases, the depth of a PDR in
pressure equilibrium with the photoionized flow were larger than the
radius of a globule, the PDR would be important in development of the
globule; if it were larger than the inter-globule separation, the
development of photoionized columns might be suppressed, with the
internal structure of the molecular cloud only observable in atomic
and molecular species.}

\subsection{Derived parameters}

Based on the observations, we will adopt the following parameters for
our modelling of column II.  The pressure in the ionized gas at the
head of the clump is $P|i/k|B \simeq 3\ee7 \cm^{-3}\Kelv$; this
ionized gas is assumed to have an isothermal sound speed of $10\kms$.
The density in the molecular gas in the core at the head is $n({\rm
H}_2) = 2\ee5\cm^{-3}$ while that in the column of gas behind is
typically 10 times smaller than this.  White \etal{} suggest that
while the temperature of the gas is $20\Kelv$ throughout the column,
the effective pressure within the columns is dominated by the effects
of small-scale magnetic fields and turbulence, as evidenced by the
smooth profiles of molecular lines with widths $\sim 2\kms$.  We cater
for this in our dynamical models by assuming that the neutral gas
behaves as if it were isothermal with a temperature of $200\Kelv$
(note that the turbulent contribution to the pressure in the molecular
gas also reduces the importance of heating in the PDR for dynamical
modelling).  The total mass in the core of column II is $31\Msun$ and
its radius is $0.085\parsec$.  We assume a photoionizing flux of
$10^{11}\cm^{-2}\secnd^{-1}$, impinging on the column parallel to its
axis.

\section{Numerical Method and Assumptions}

\label{s:num}
We use the hydrodynamical code described by Williams
\shortcite{will99}, which is based on the second-order Godunov
algorithm by Falle~\shortcite{falle}.  Ionization and recombination are
treated by including the equations
\begin{eqnarray}
n\rdif{x}{t} &=& a n (1-x) J - \alpha|B n^2 x^2 \\
\rdif{J}{z} &=& -a n (1-x) J. 
\end{eqnarray}
Here $n$ is the density of hydrogen nucleons, $x$ the ionization
fraction and $J$ is the number flux of ionizing photons, assumed to
propagate parallel to the $z$ axis.  The photoionization cross-section
of hydrogen, $a$, and the case B recombination coefficient,
$\alpha|B$, are taken as constants.  The electron density, $nx$, is
included as an advected variable in the hydrodynamic solver, and the
ionization and recombination equations included by operator-splitting,
implemented at the end of each step (and also for the intervening
half-steps).  The equations are integrated using an implicit scheme to
advance $x$ in time and $J$ across each grid cell.

The code does not include a treatment of the diffuse ionizing
radiation field (beyond using the case-B recombination coefficient).
In the on-the-spot approximation, the diffuse field is found to be
$\sim 15$ per cent of the {\it local}\/ direct radiation field
\cite{cantea98}.  The characteristic length for reabsorption of
diffuse photons is $\ell\sim J/\alpha|B n^2 \simeq 0.05\parsec$
(dependent, or course, on spatial variations of $J$ and $n$), so the
diffuse field might be expected to have some effect even on structures
as large as the columns.  More detailed treatments predict that the
diffuse radiation is a smaller fraction of the direct field away from
the edge of the global photoionized region \cite{rubin68}.  Close to
the edge, where relative size of the diffuse field is larger, the soft
spectrum of the diffuse field will result in a smaller path length to
absorption than for the direct field.  The low contrast of the
surfaces of the observed columns which do not point towards the
ionizing source is empirical evidence that the diffuse radiation field
would is adequately treated by the on-the-spot approximation in the
present case.

Since the internal structures of IFs are not resolved in the
simulations, most of the gas will be either almost completely ionized
or fully neutral.  Intervening regions with intermediate ionization
are likely to result principally from numerical mixing, so physically
detailed schemes which assume the gas within the computational cells
is reasonably uniform are not appropriate.  Instead, we set isothermal
sound speed in the gas to
\begin{equation}
c|s = (1 + 99 x)^{1/2} \kms.\label{e:cs}
\end{equation}
This expression results from assuming that in cells with $0<x<1$, the
gas is split into separate regions of fully ionized and fully neutral
gas, in pressure equilibrium with each other, and so gives a better
account of the sub-grid structure at the (dynamically significant)
D-type IF.

\section{Initial and boundary conditions}

\label{s:cond}
The main model presented here is calculated in cylindrical symmetry,
with coordinates $(r,z)$.  A uniform $128\times448$ grid covers
physical dimensions $0<r<0.5\parsec$ and $0<z<1.75\parsec$.  The
boundary conditions are reflective on the axis, $r=0$, as required by
symmetry.  On the boundary towards the source, $z=1.75\parsec$,
free-flow boundary conditions are used when the flow is outwards,
mirror conditions when it is inwards, although as the outward flow is
generally supersonic this has little influence on the flow inside the
grid.  Along the outer radial boundary, $r=0.5\parsec$, the conditions
beyond the grid are either free-outflow/zero-inflow if the ionization
fraction is larger than 0.1, or mirror conditions if it is less than
this value.  This somewhat artificial criterion changes the behaviour
at the boundary at the IF, in order to model an ionized flow which is
divergent on large scales.  The effect depends little on the specific
limiting value, as away from the IF the gas is in general either
almost entirely ionized or almost entirely neutral.  

On the edge of the grid towards the molecular cloud, $z=0$, the
boundary condition is set by assuming that the material retains its
initial state, i.e.\ in most of our examples a flow of gas towards the
IF\@.  As the solution develops, a shock often becomes captured at the
boundary, which reduces the inflow speed and increases the inflow
density.  The flux of neutral gas into the grid through the plane
$z=0$ is, in all cases, small enough that it could not prevent the IF
from entering if the entire grid were full of neutral material.  In
many cases, the density of the ionized gas becomes large enough to
absorb all the ionizing photons within the size of the grid.  Where
the `molecular' boundary covers the entire upstream edge of the grid,
a plane-parallel IF is an equilibrium solution.  If the ionized flow
is allowed to escape from the side of the region, it is not obvious
what the form of the equilibrium will be, if indeed there is one.  One
possibility is that the IF surface will remain smooth, and a steady
solution will be found in which the variation in pressure at the IF is
balanced by changes in flow velocity via the Bernoulli effect.

However, these IF are believed to be unstable to corrugation modes on
both small and large scales, although for wavelengths larger than
$6\ee{15}n_3^{-1}\cm$ recombination decreases the growth rate of these
instabilities decreases rapidly \cite{kahn58,axfo64,sysoev97}.  This
wavelength is comparable to our grid scale of $10^{16}\cm$ for typical
ionized gas densities at the IF of $\sim 10^{3}\cm^{-3}$.  We now
present numerical models which illustrate what may happen for a range
of initial conditions.

The models we have calculated have initial conditions split into two
regions: ionized gas with a density $n({\rm H\,total}) = 200\cm^{-3}$
towards the source of ionizing photons and moving towards it at
$10\kms$ relative to the neutral gas behind.  This neutral gas has
density, $n({\rm H\,total}) = 2\ee4\cm^{-3}$ (n.b., not $n({\rm
H}_2)$, see below).  Except where noted, in the initial conditions the
neutral gas is assumed to move at $2\kms$ towards the ionizing source
in order to keep the ionization front broadly at rest within the grid,
with the ionized gas assumed to stream towards the radiation source
$10\kms$ more rapidly than this.  Pressure equilibrium holds where
ionized and neutral gas are alongside each other.  We take the
ionizing flux incident at the top of the grid in the figures to be
$1.5\ee{11}\cm^{-2}\secnd^{-1}$, since recombinations in the upstream
flow are found to reduce the ionizing radiation incident on the head
of the column somewhat.  The Str\"omgren distance in the ionized gas,
$\sim 3\parsec$, is sufficiently great that a column the length of the
computational grid can be maintained with almost complete ionization
for the initial conditions.

The boundary between these regions is assumed either to be plane (to
study the development for small perturbations) or to include a column
of dense gas (to study the likely long-timescale development of the
columns observed in many regions).

\section{Results}

We will now present the results of four different scenarios for the
development of photoevaporated columns. 
\begin{description}
\item[{\it Case I}] is based on the model of White
\etal~\shortcite{white99}, including initial density stratification
and gravity,
\item[{\it Case II}] has an initial short cylindrical column,
\item[{\it Case III}] has a 10 per cent deficit in photoionizing flux
close to the axis,
\item[{\it Case IV}] has dense neutral gas initially filling the grid,
with gravitational forces from a $30\Msun$ object included.
\end{description}

\begin{figure*}
\begin{centering}
\begin{tabular}{lll}
(a) & (b) & (c) \\
\epsfysize=5cm\rotatebox{270}{\epsffile{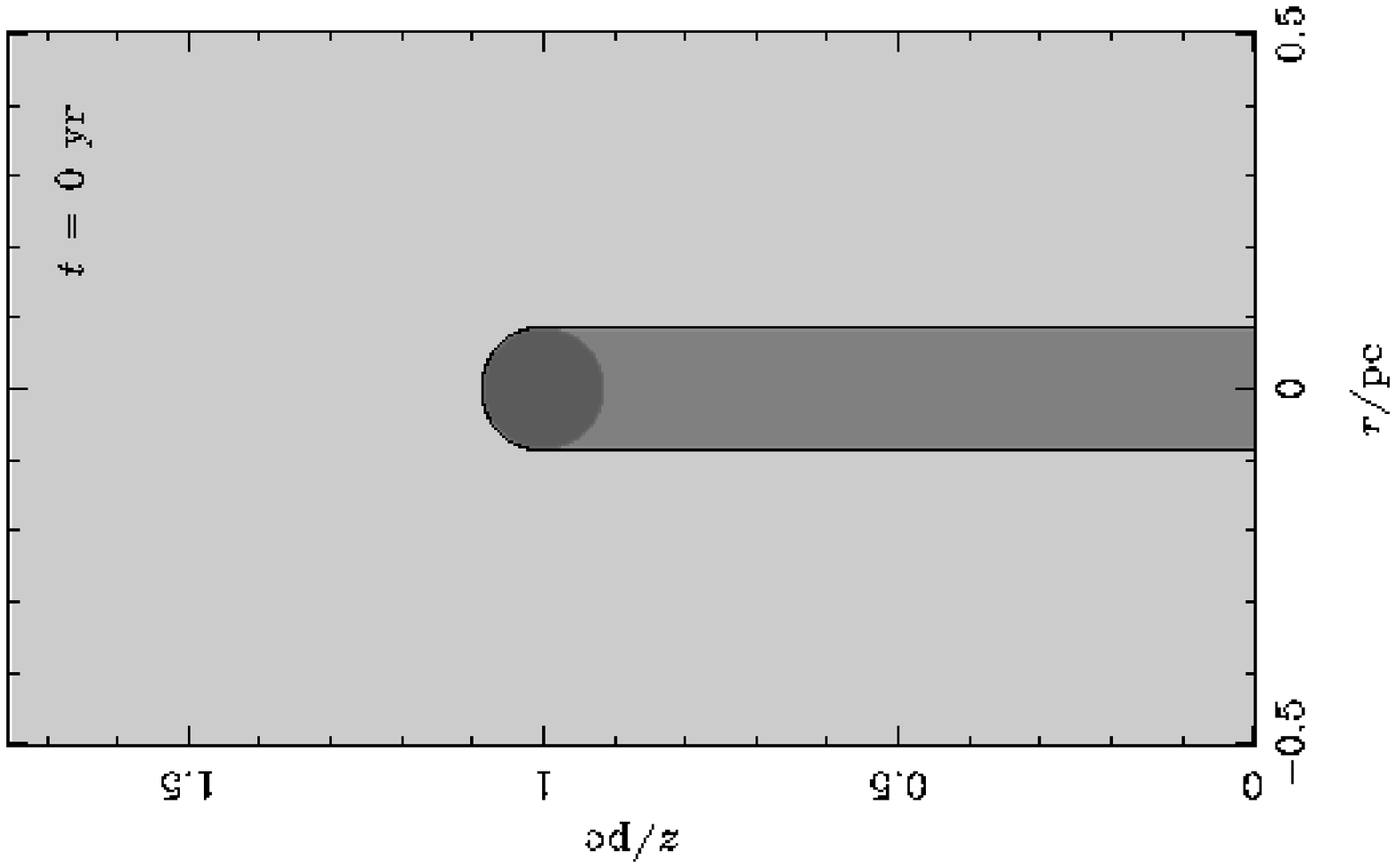}} &
\epsfysize=5cm\rotatebox{270}{\epsffile{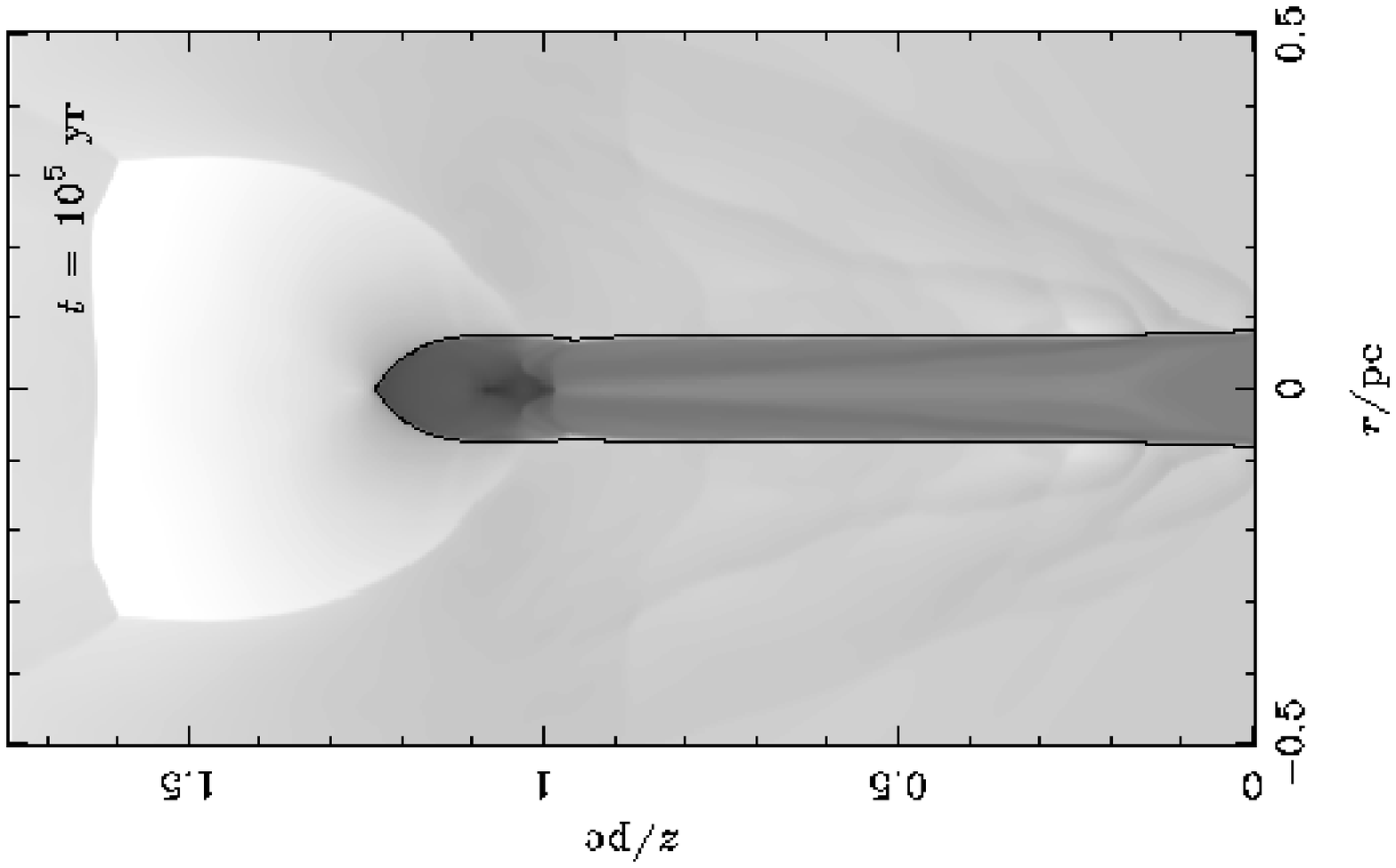}} &
\epsfysize=5cm\rotatebox{270}{\epsffile{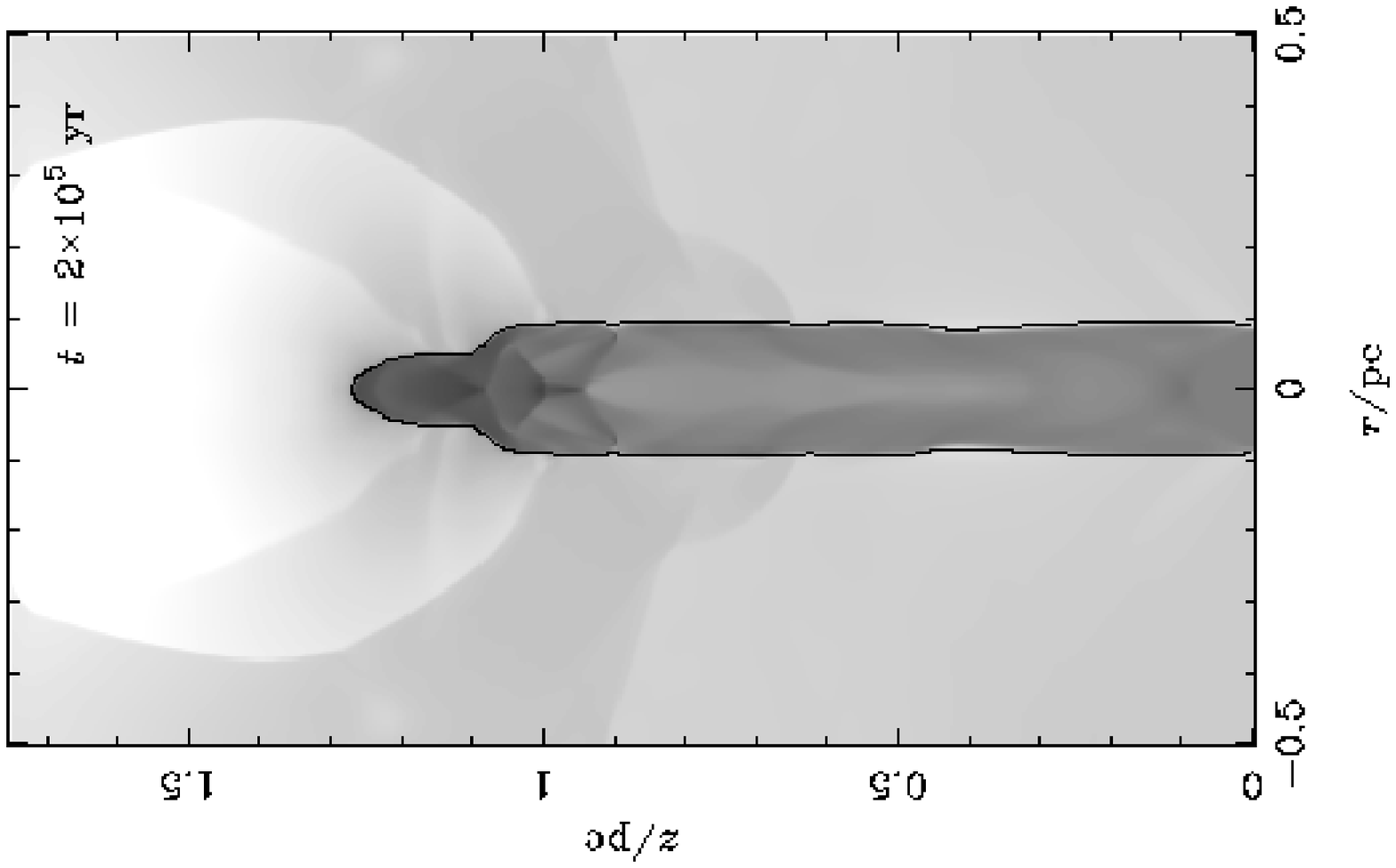}} \\
(d) & (e) & (f) \\
\epsfysize=5cm\rotatebox{270}{\epsffile{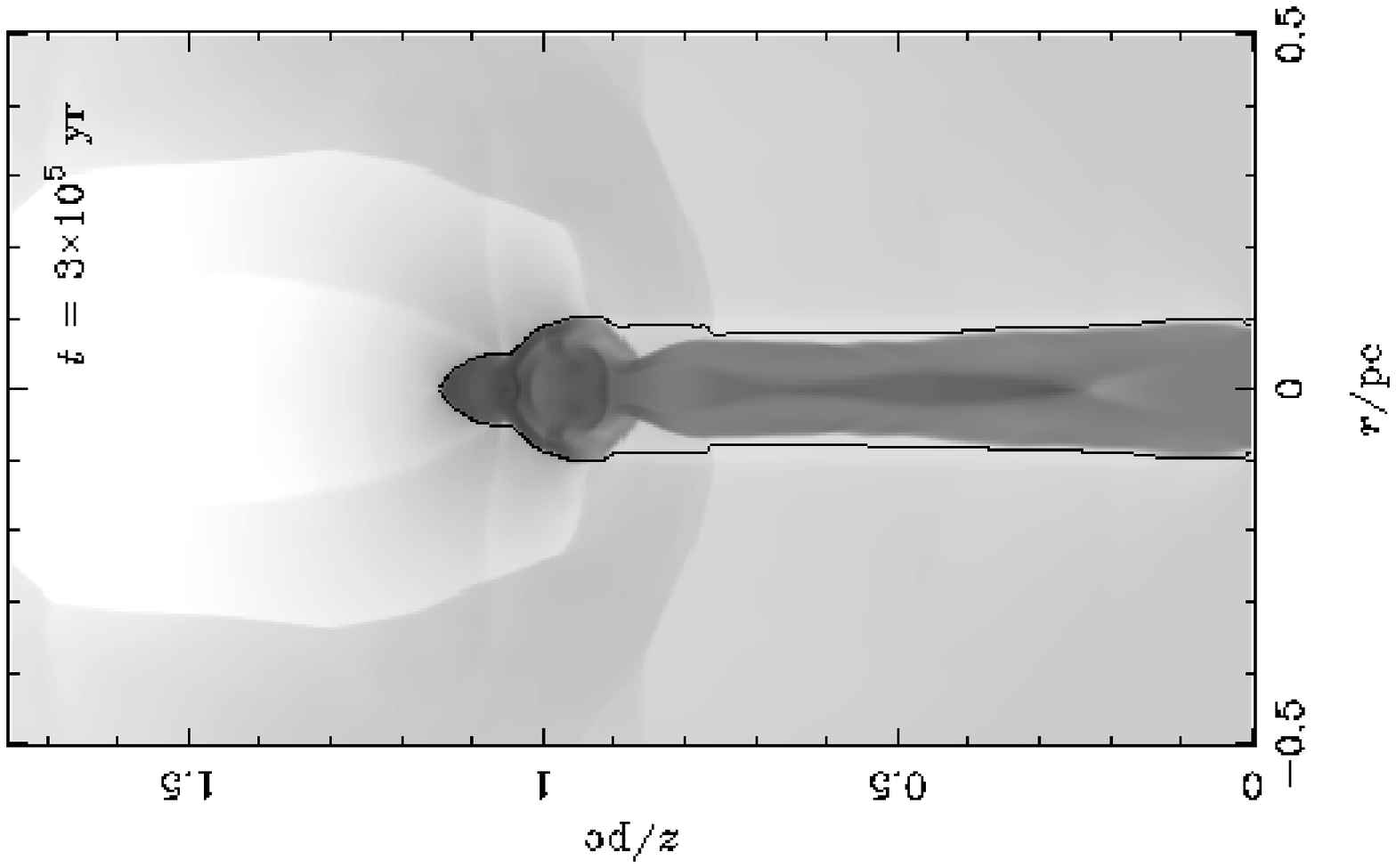}} &
\epsfysize=5cm\rotatebox{270}{\epsffile{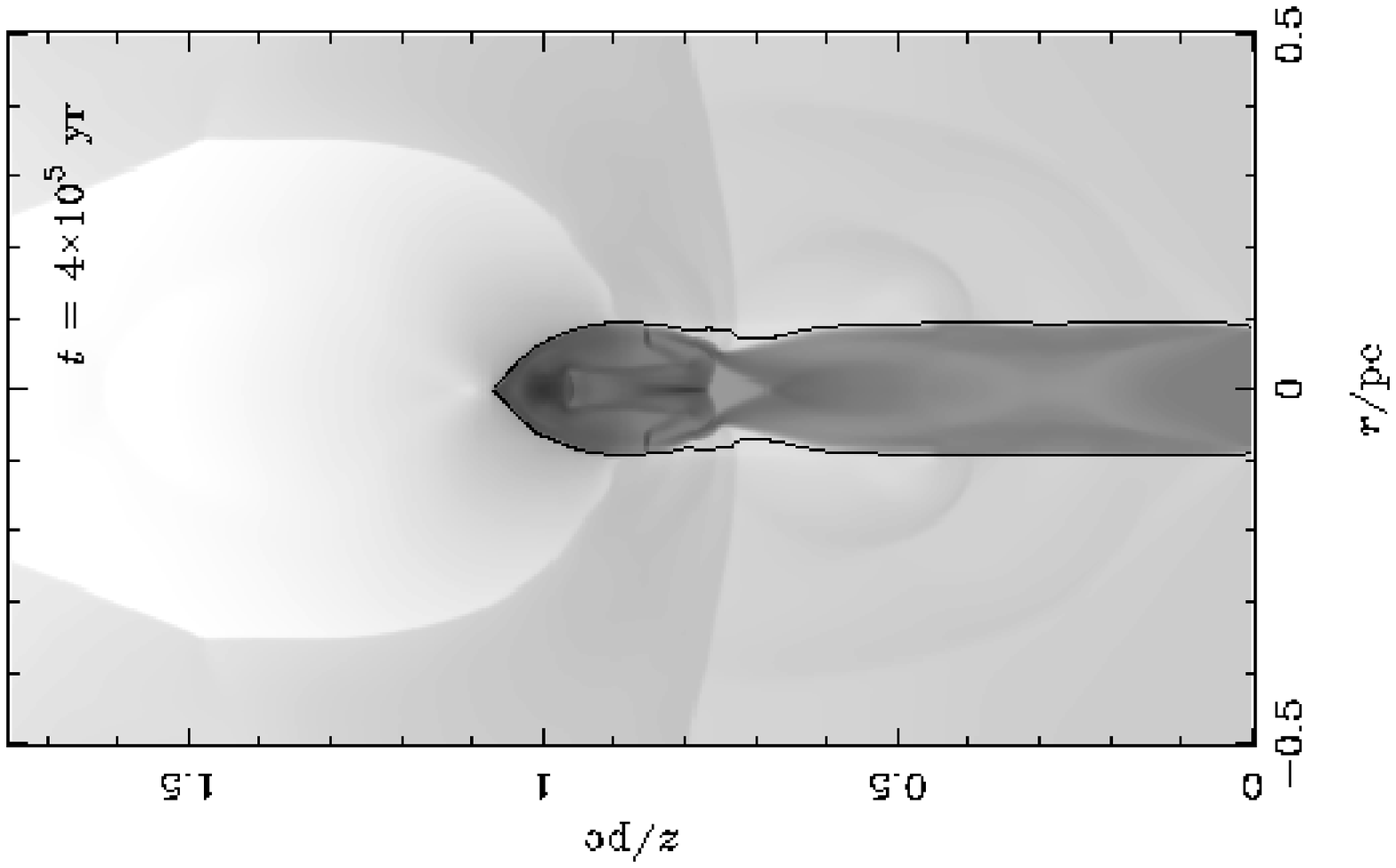}} &
\epsfysize=5cm\rotatebox{270}{\epsffile{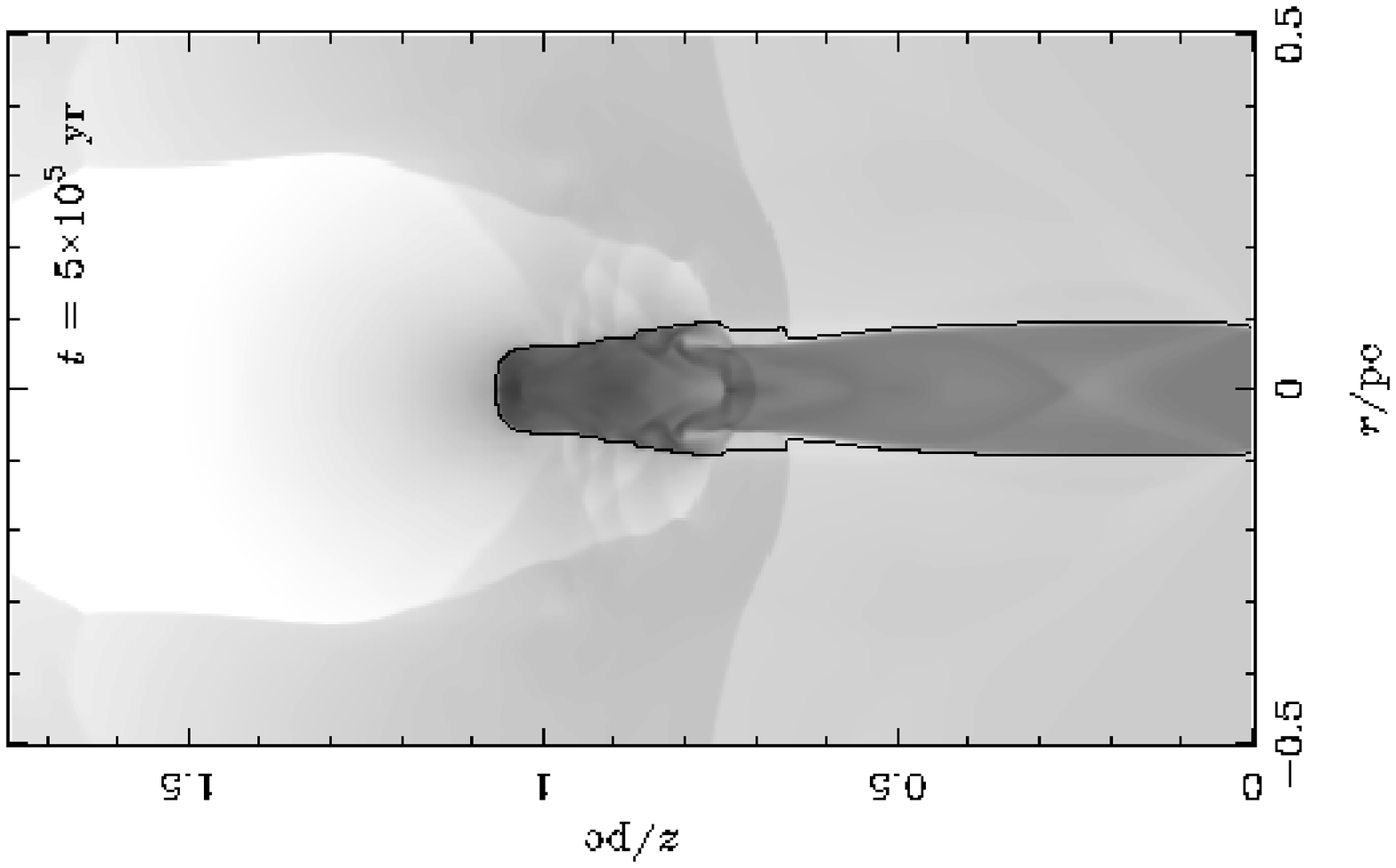}} \\
\multicolumn{3}{c}{\epsfysize=10cm\rotatebox{270}{\epsffile{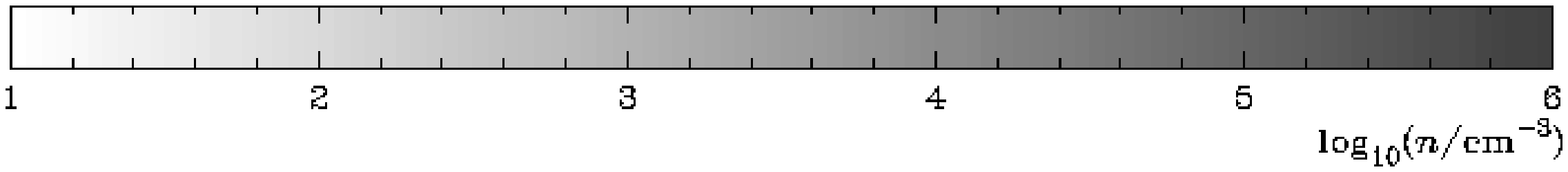}}}
\end{tabular}
\end{centering}
\caption{Greyscale plots of log density for Case I: an initial column
of gas based on the model of White \etal{} The plots are shown at the
labelled times after the start of the simulation.  The density range
calibration is shown in the colour bar; a single contour marks the
location of the ionization front.}
\label{f:simwhite}
\end{figure*}

\subparagraph{Case I:}

In Figure~\ref{f:simwhite}, we present the results of a first
simulation in which the flow conditions modelled on those inferred by
White \etal{} The initial conditions have a dense core of material at
the head, and a lower density column behind.  A gravitational
potential from a $30\Msun$ mass, smoothed at a radius of
$0.08\parsec$, is centred on the core of the head.  The uniform gas at
the head of the column initially falls in, as it is not in equilibrium
with the gravitational field, but does not collapse far because of the
smoothed core of the field.

By the time of the second frame, $10^5\yr$ after the start of the
simulation, the shock at the head of the the column has passed through
much of the dense gas.  The shock proceeds down the column in later
frames.  Soon the ram pressure of the radiation field is sufficient to
push the dense gas away from the gravitating core.  The amount of gas
with density above $10^5\cm^{-3}$ in these models remains roughly
constant, and so the gravitational field would in reality respond to
this movement.  Nevertheless, it is striking that the density field in
the column remains close to that inferred by White \etal{} even when
the head is a considerable distance away from the gravitating centre,
rather than requiring the head of the column to be caught within
$10^5\yr$ after the IF reaches it.

In the remaining examples, we investigate possibilities by which such
structures may have formed, and what conditions are necessary for them
to be long-lived.

\label{s:res}
\begin{figure*}
\begin{centering}
\begin{tabular}{lll}
(a) & (b) & (c) \\
\epsfysize=5cm\rotatebox{270}{\epsffile{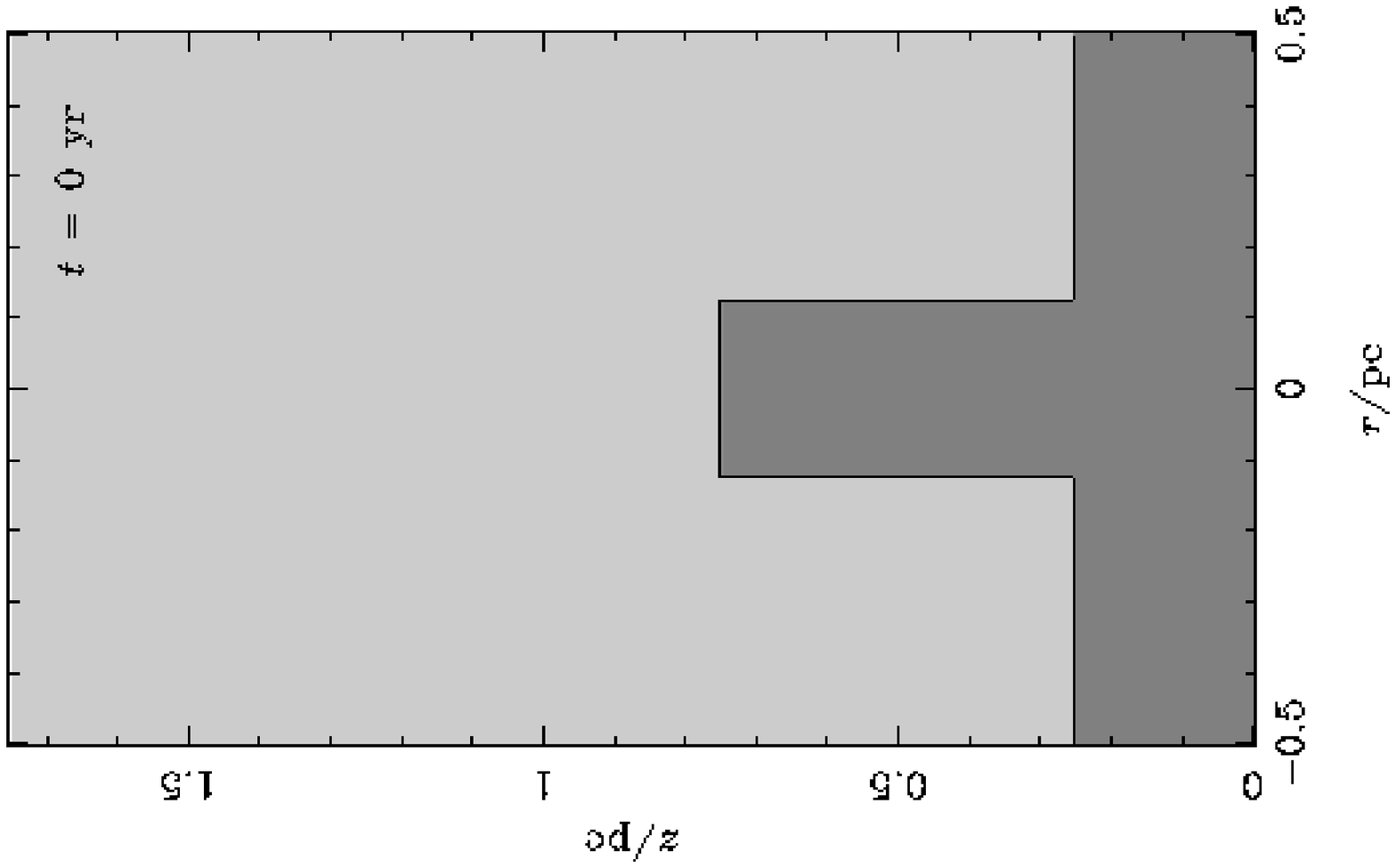}} &
\epsfysize=5cm\rotatebox{270}{\epsffile{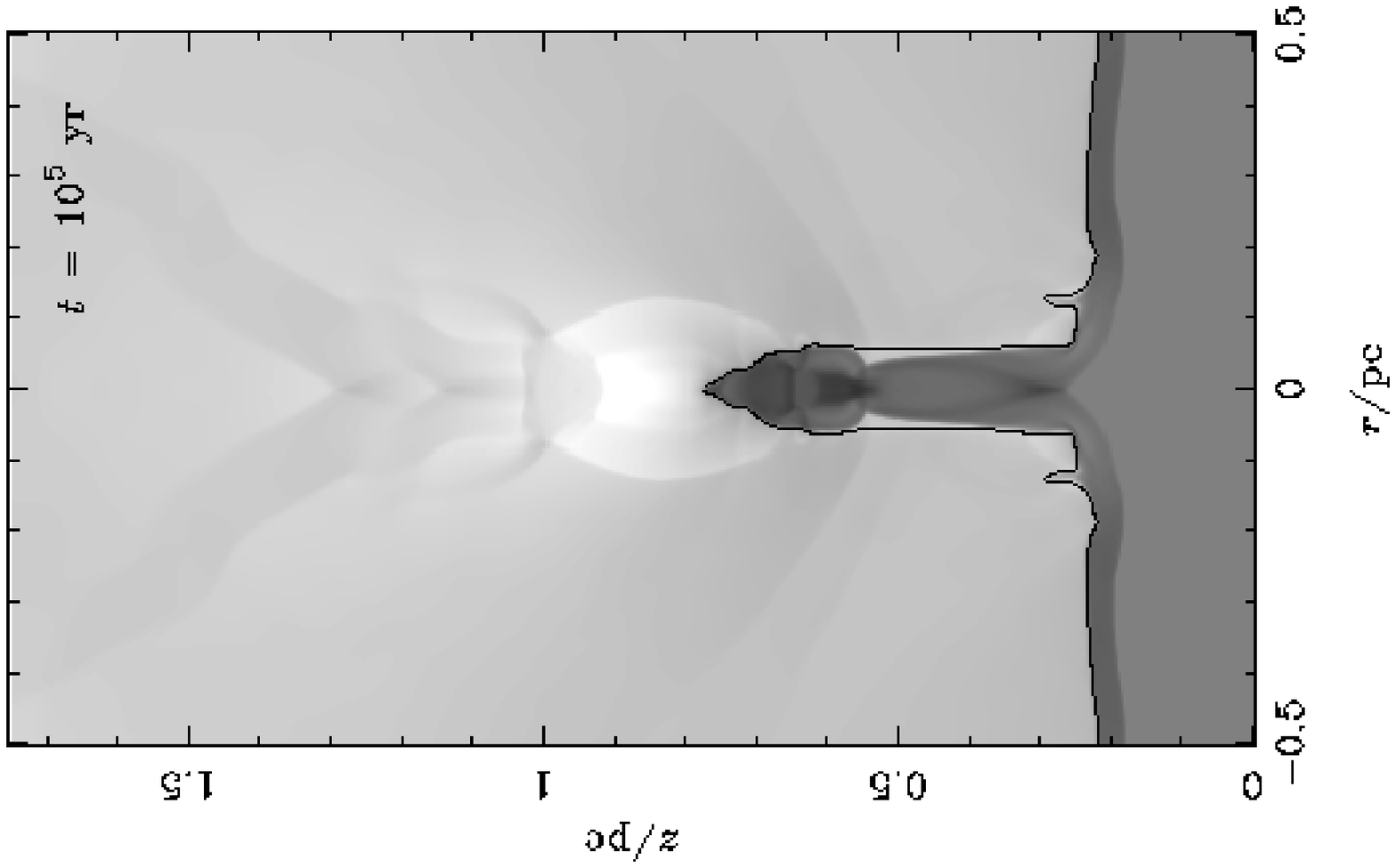}} &
\epsfysize=5cm\rotatebox{270}{\epsffile{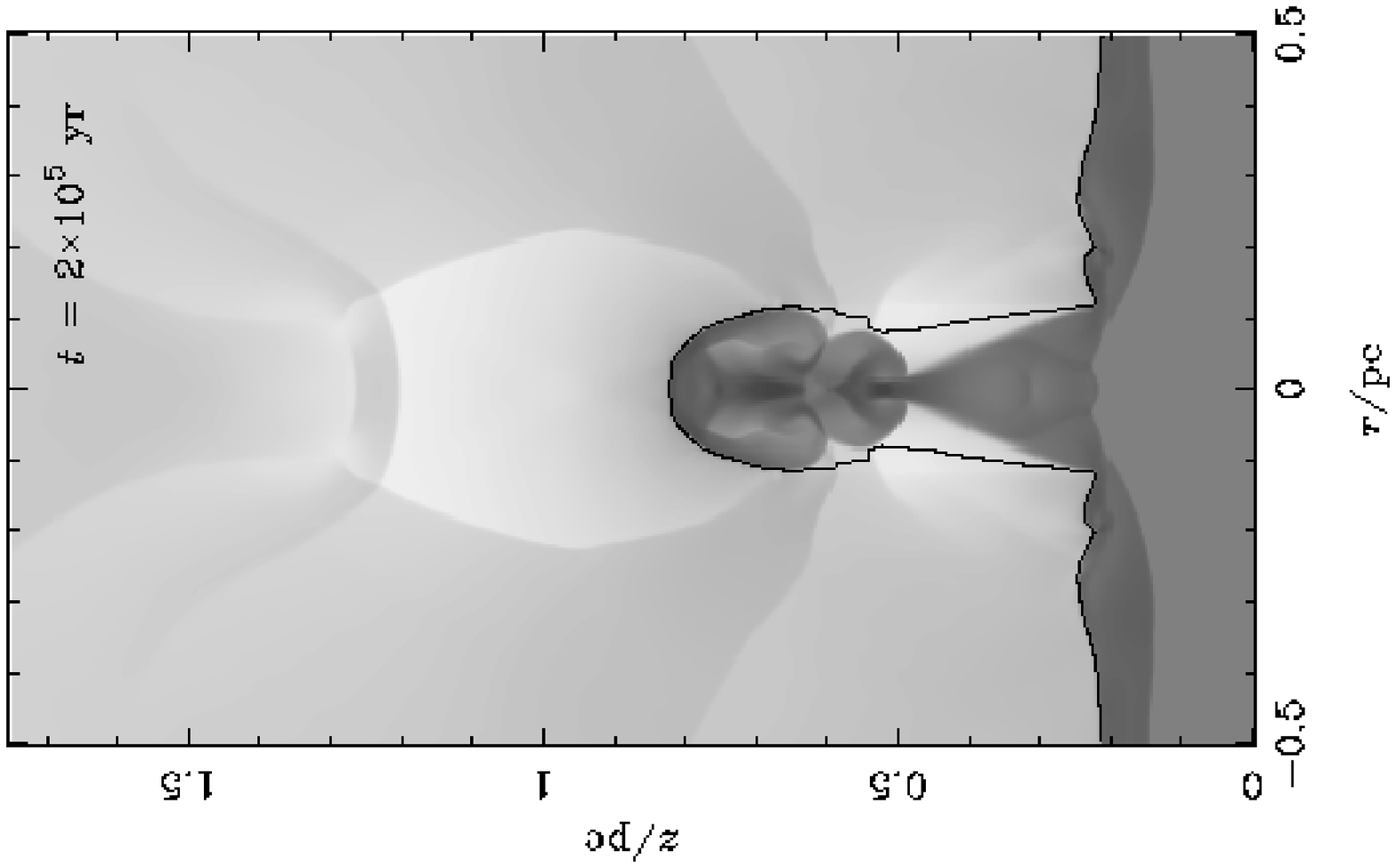}} \\
(d) & (e) & (f) \\
\epsfysize=5cm\rotatebox{270}{\epsffile{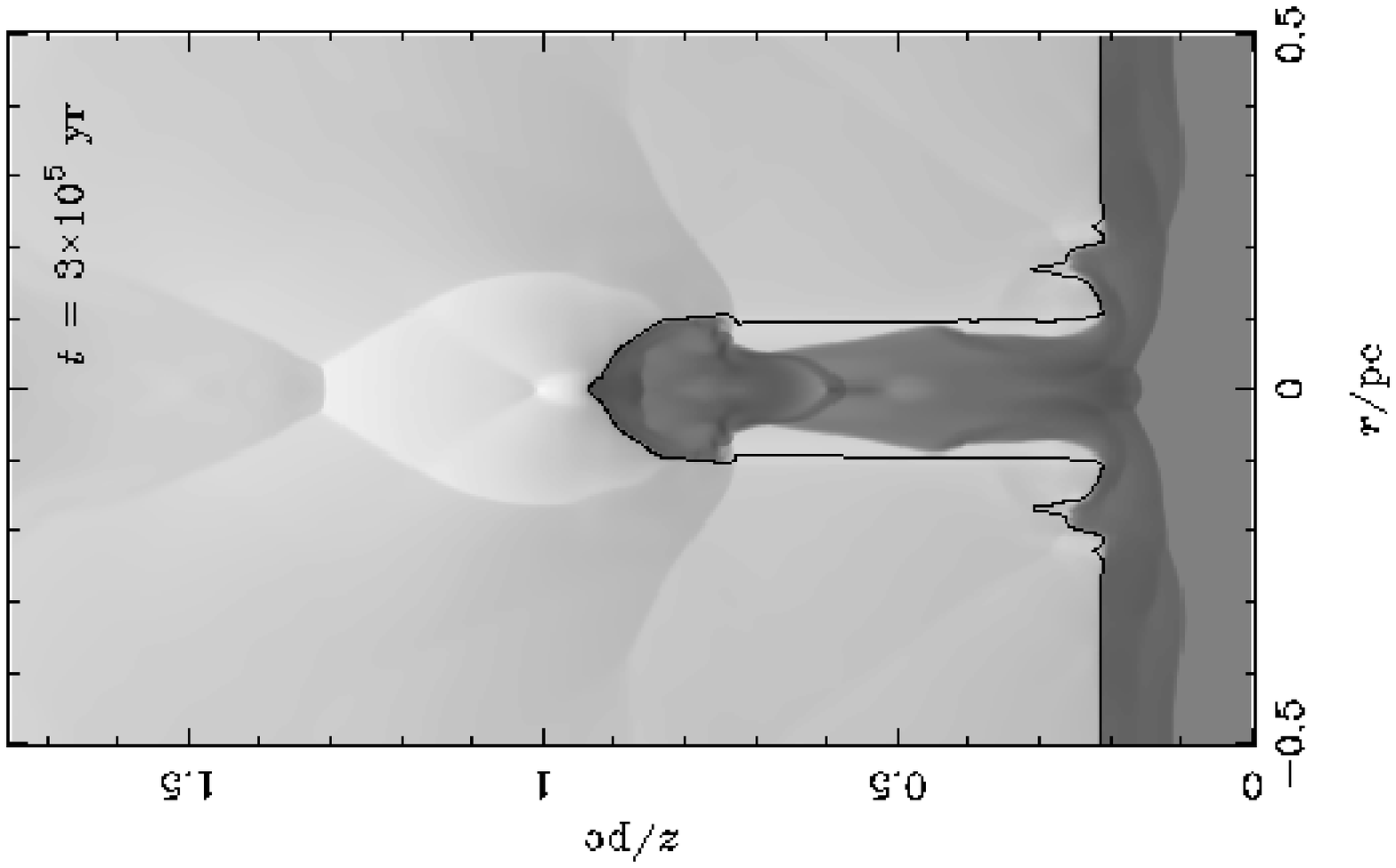}} &
\epsfysize=5cm\rotatebox{270}{\epsffile{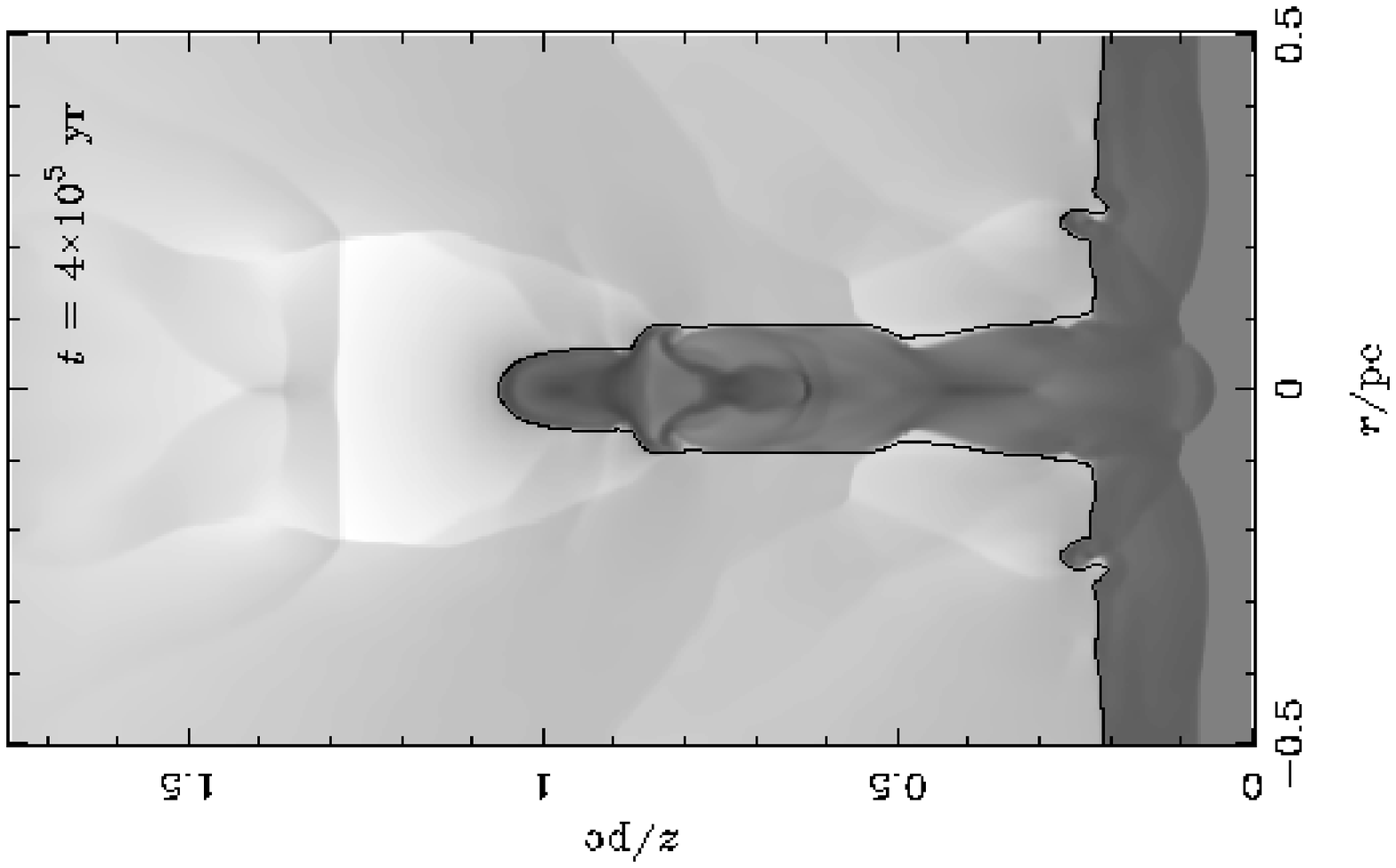}} &
\epsfysize=5cm\rotatebox{270}{\epsffile{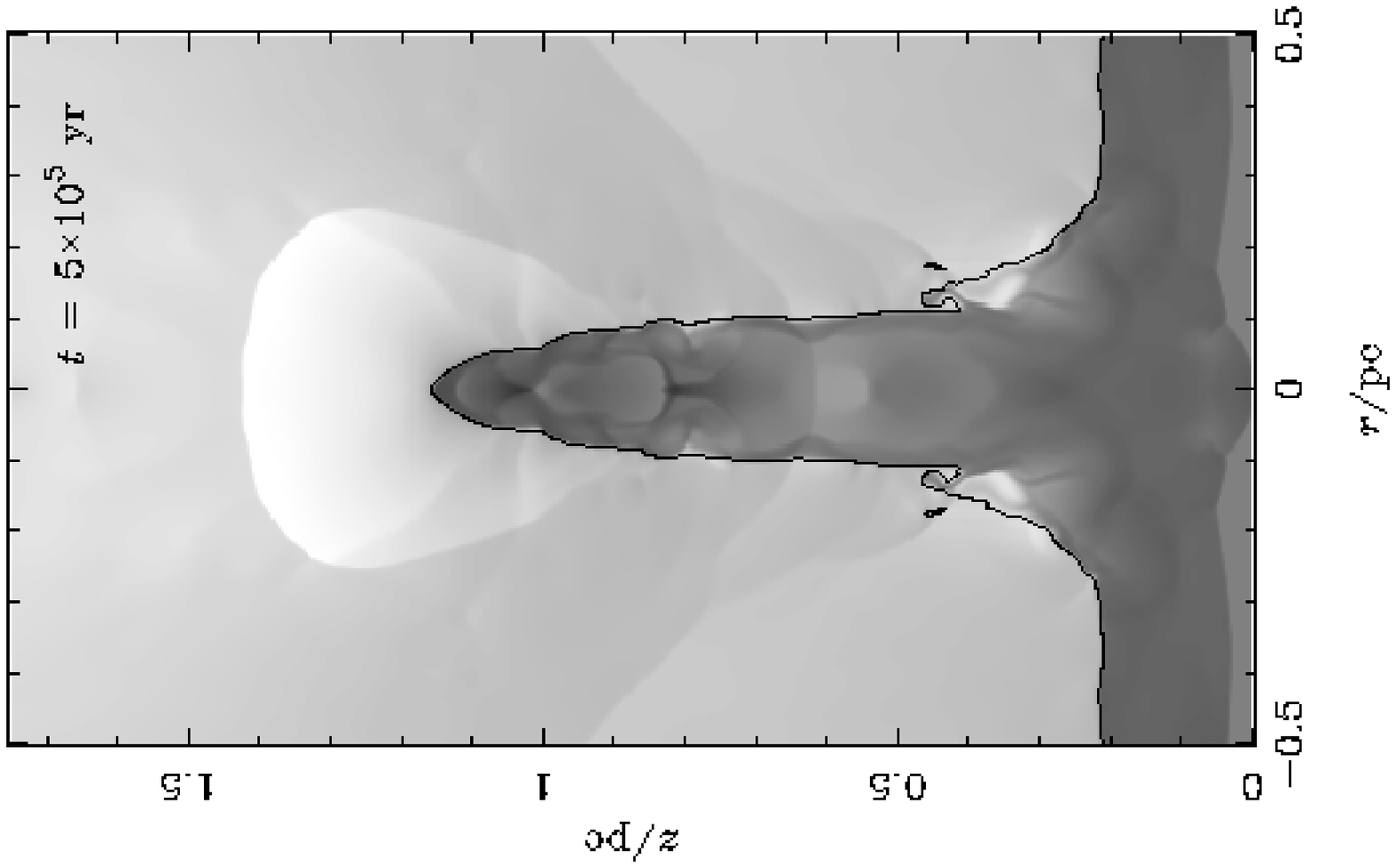}}\\
\multicolumn{3}{c}{\epsfysize=10cm\rotatebox{270}{\epsffile{wedge.ps}}}
\end{tabular}
\end{centering}
\caption{Greyscale plots of log density for Case II.  The plots are
shown at (a) initial time (b) $10^5\yr$, (c) $2\ee5\yr$, (d)
$3\ee5\yr$, (e) $4\ee5\yr$, and (f) $5\ee5\yr$ after the start of the
simulation.  The full density range is typically from $3$ to
$10^6\cm^{-3}$, although the gas which fills most of the volume of the
column is between $5\ee4$ and $5\ee5\cm^{-3}$.}
\label{f:simdensb}
\end{figure*}

\subparagraph{Case II:}

In Figure~\ref{f:simdensb}, we illustrate a case in which a short
column of gas is initially present.  The radius of the column is
initially $0.125\parsec$ and it stands proud by $0.5\parsec$ ahead of
the main body of molecular material.  The column is again seen to be
long lived, but highly variable.  Certain broad features characterise
the structure.  The highest densities in the neutral gas are seen
close to the head of the column, and reach values similar to those
inferred for the column tip by White \etal{}

The surface of the column is convoluted.  Neutral clumps split off
from from the column at different stages (\eg{} close to the base of
the column in Figure~\ref{f:simdensb}f) and linear structures are seen
where different streams of ionized gas interact.  These are
reminiscent of the observed clumps and striations in the flow (see
Figure~\ref{f:image} and Section~\ref{s:deriv}).  The striations are
sometimes seen to be triggered by surface inhomogeneities resulting
from the discrete nature of the grid.  However, the growth of these
instabilities appears to be a generic feature of the development of
obliquely-illuminated ionization fronts \cite{van62}, even when
account is taken of the effects of recombination in the upstream flow
(Williams, in preparation).

In the frames shown in Figure~\ref{f:simdensb}, the dense column
becomes more marked with time.  Initially, the ionizing field
generates a strong flow away from the head of the column, a low
density region ahead of it and a net acceleration away from the
ionizing source \cite[the well-known `rocket effect',]{os54}.
However, we see here that short lengthscale which characterises the
photoevaporative flow from the head means that its pressure decreases
rapidly, and eventually a termination shock forms.  The convergence of
this termination shock at the axis leads to the formation of an
over-dense region at larger distances from the column.  Such
over-dense regions lead to the formation of some thin filaments
connected to the main ionization front [e.g., those seen close to the
base of the column in Figure~\ref{f:simdensb}(b)].  Filaments at the
head of the column can allow the column to move forward once again,
although the rigidity provided for the filaments by the reflecting
boundary condition on-axis may have a significant influence by
preventing non-axisymmetric instability modes.

\begin{figure*}
\begin{centering}
\begin{tabular}{lll}
(a) & (b) & (c) \\
\epsfysize=5cm\rotatebox{270}{\epsffile{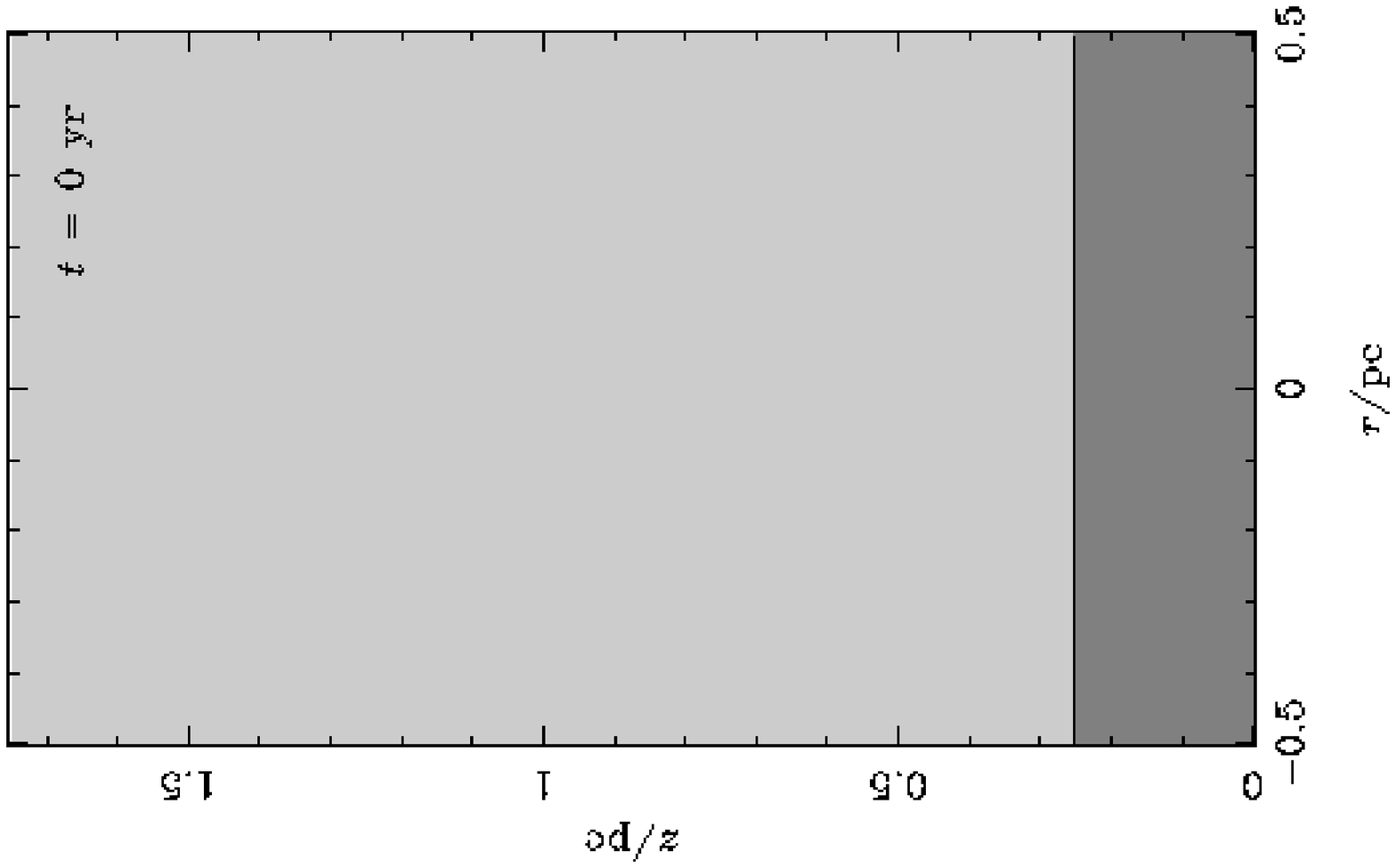}} &
\epsfysize=5cm\rotatebox{270}{\epsffile{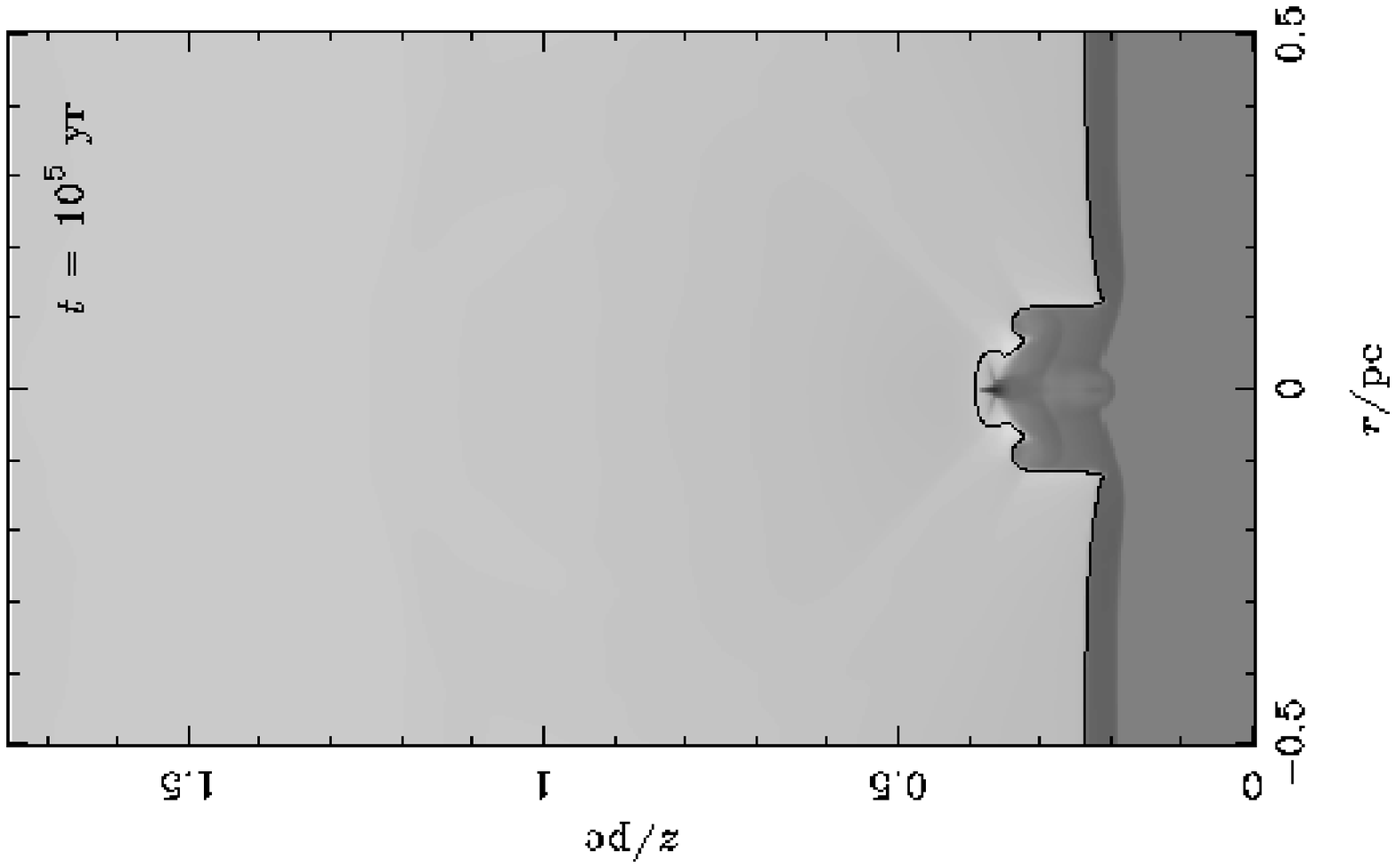}} &
\epsfysize=5cm\rotatebox{270}{\epsffile{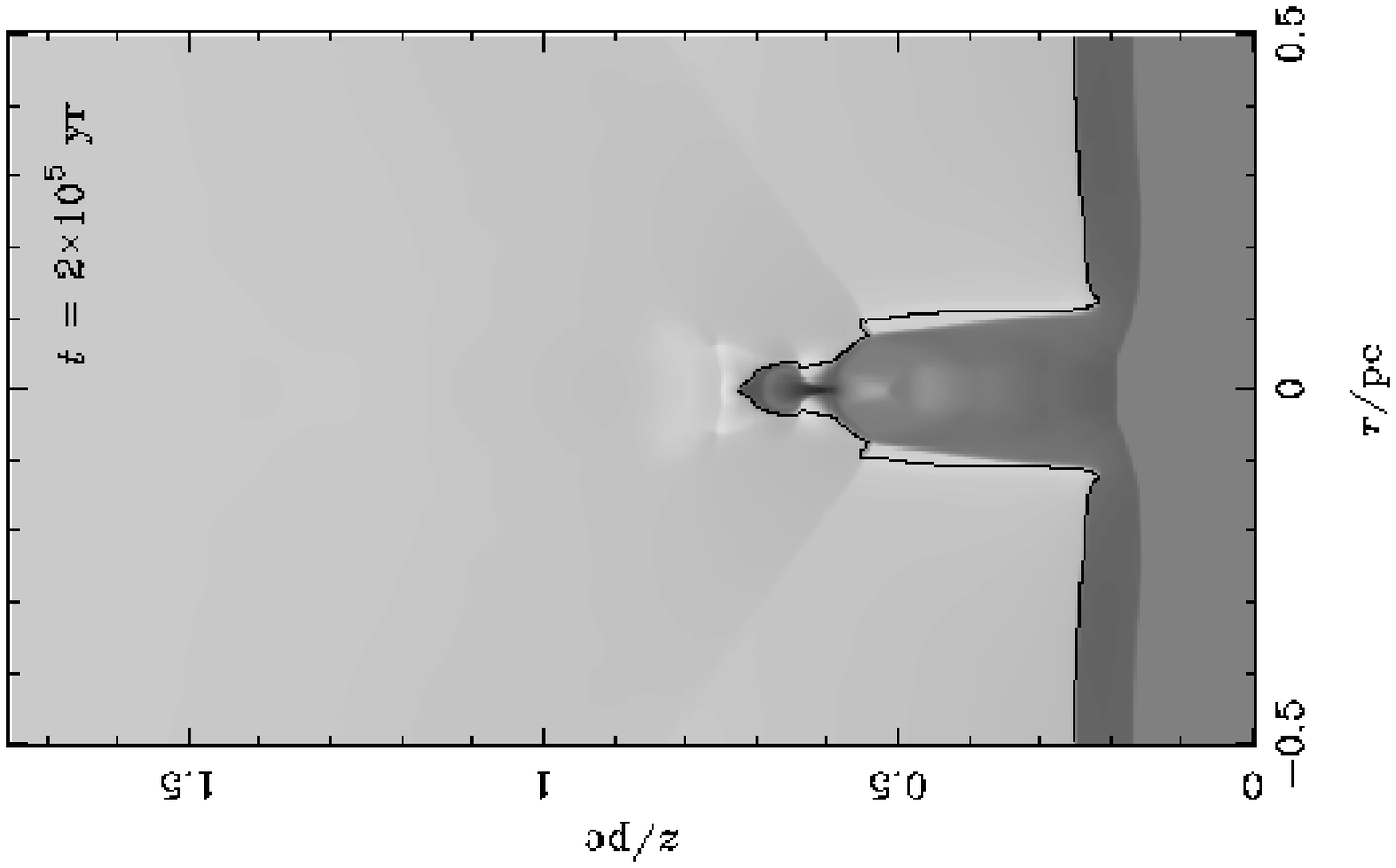}} \\
(d) & (e) & (f) \\
\epsfysize=5cm\rotatebox{270}{\epsffile{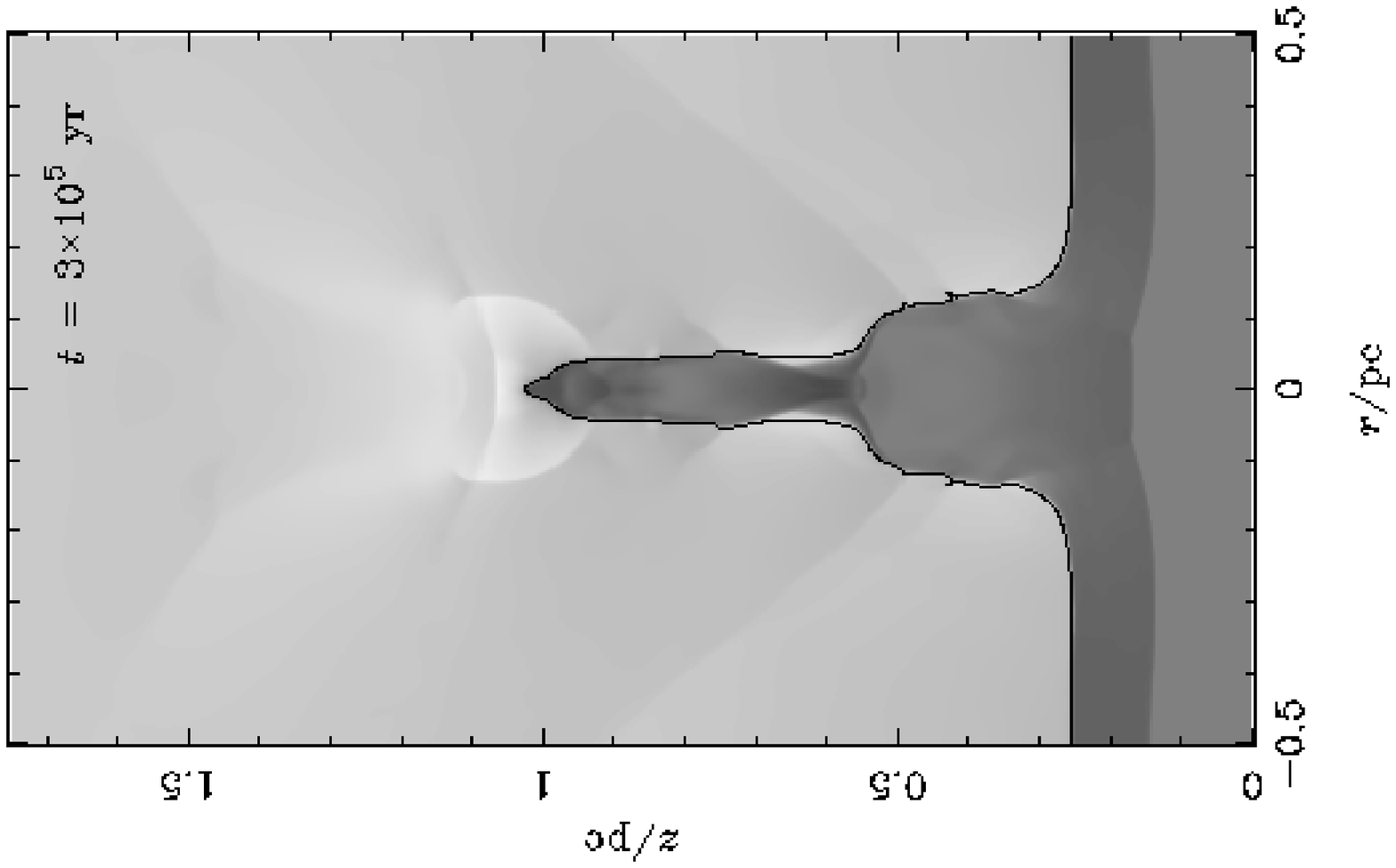}} &
\epsfysize=5cm\rotatebox{270}{\epsffile{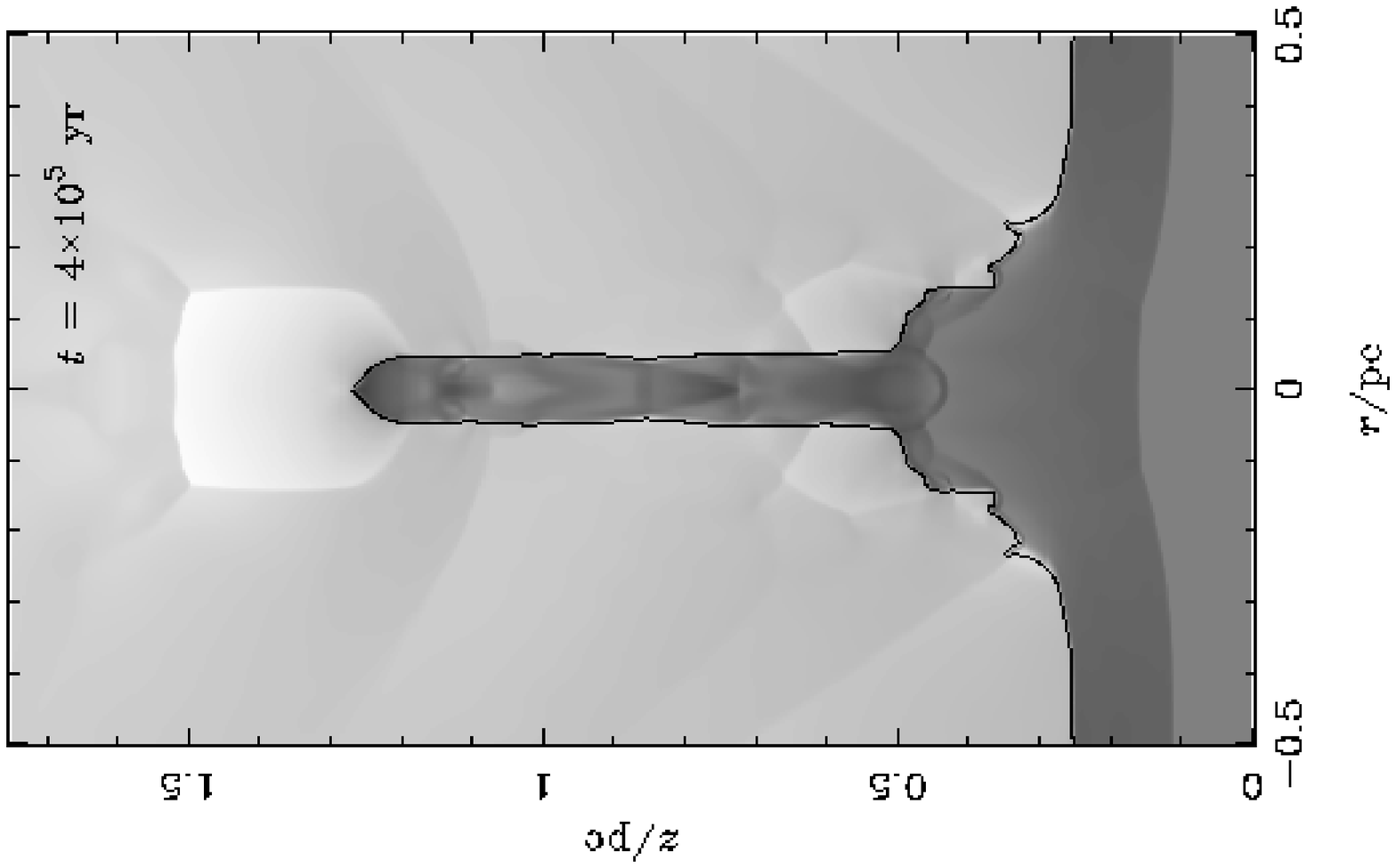}} &
\epsfysize=5cm\rotatebox{270}{\epsffile{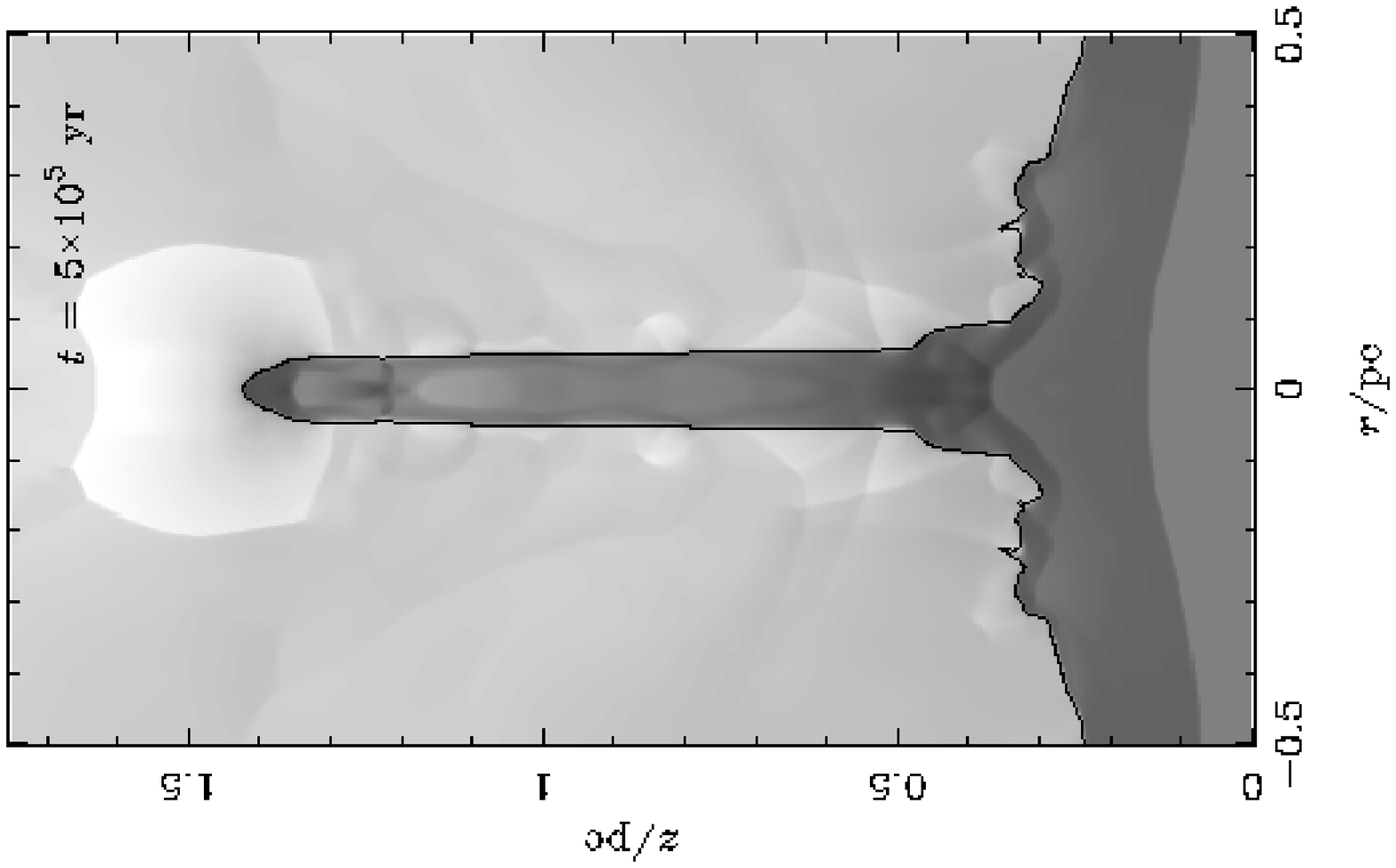}}\\
\multicolumn{3}{c}{\epsfysize=10cm\rotatebox{270}{\epsffile{wedge.ps}}}
\end{tabular}
\end{centering}
\caption{Greyscale plots of log density for Case III, stationary,
initially uniform gas with 10 per cent decrease in ionizing radiation
field within a $0.125\parsec$ radius of the axis.  The plots are shown
at (a) initial conditions, (b) $1\ee5\yr$, (c) $2\ee5\yr$, (d)
$3\ee5\yr$, (e) $4\ee5\yr$, and (f) $5\ee5\yr$ after the start of the
simulation.  The density range calibration is shown in the colour
bar.}
\label{f:simshad}
\end{figure*}

\subparagraph{Case III:}

In Figure~\ref{f:simshad}, we show the development of columns for a
case where there is a 10 per cent weakening of the impinging ionizing
radiation field for a small radius around the axis.  A narrow feature
in the radiation field such as this might formed by an upstream
overdensity: a photoevaporating globule, for instance, or dense
structures close to the stars caused by interacting stellar winds.

A strong column soon forms in this case.  Once it has formed, its
evolution is not greatly influenced by the small perturbation in the
radiation field.  The structure is broadly similar to those seen in
the previous two cases, although the column has a somewhat higher
aspect ratio and the shock remains within the computational grid for a
longer period.

\begin{figure*}
\begin{centering}
\begin{tabular}{lll}
(a) & (b) & (c) \\
\epsfysize=5cm\rotatebox{270}{\epsffile{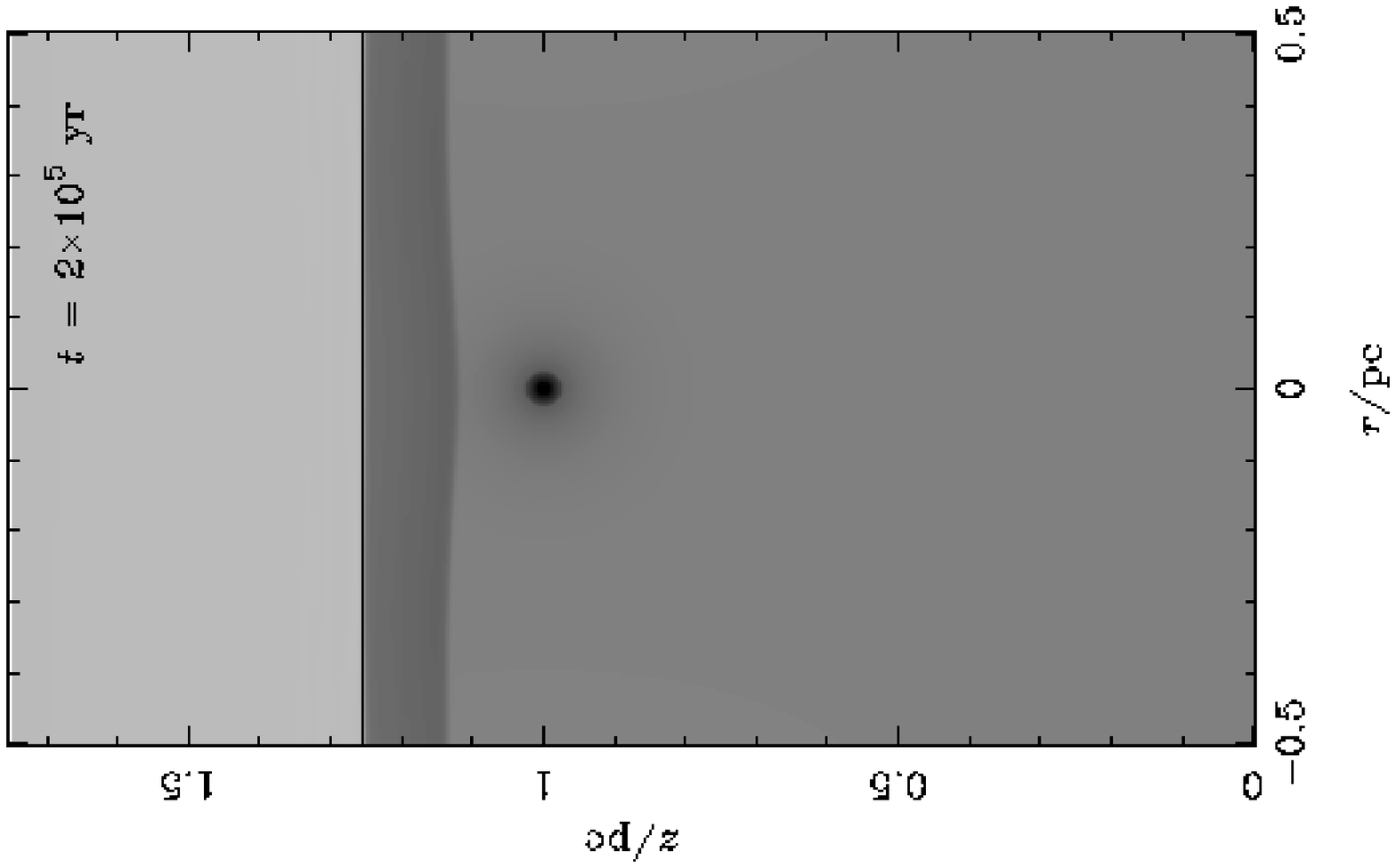}} &
\epsfysize=5cm\rotatebox{270}{\epsffile{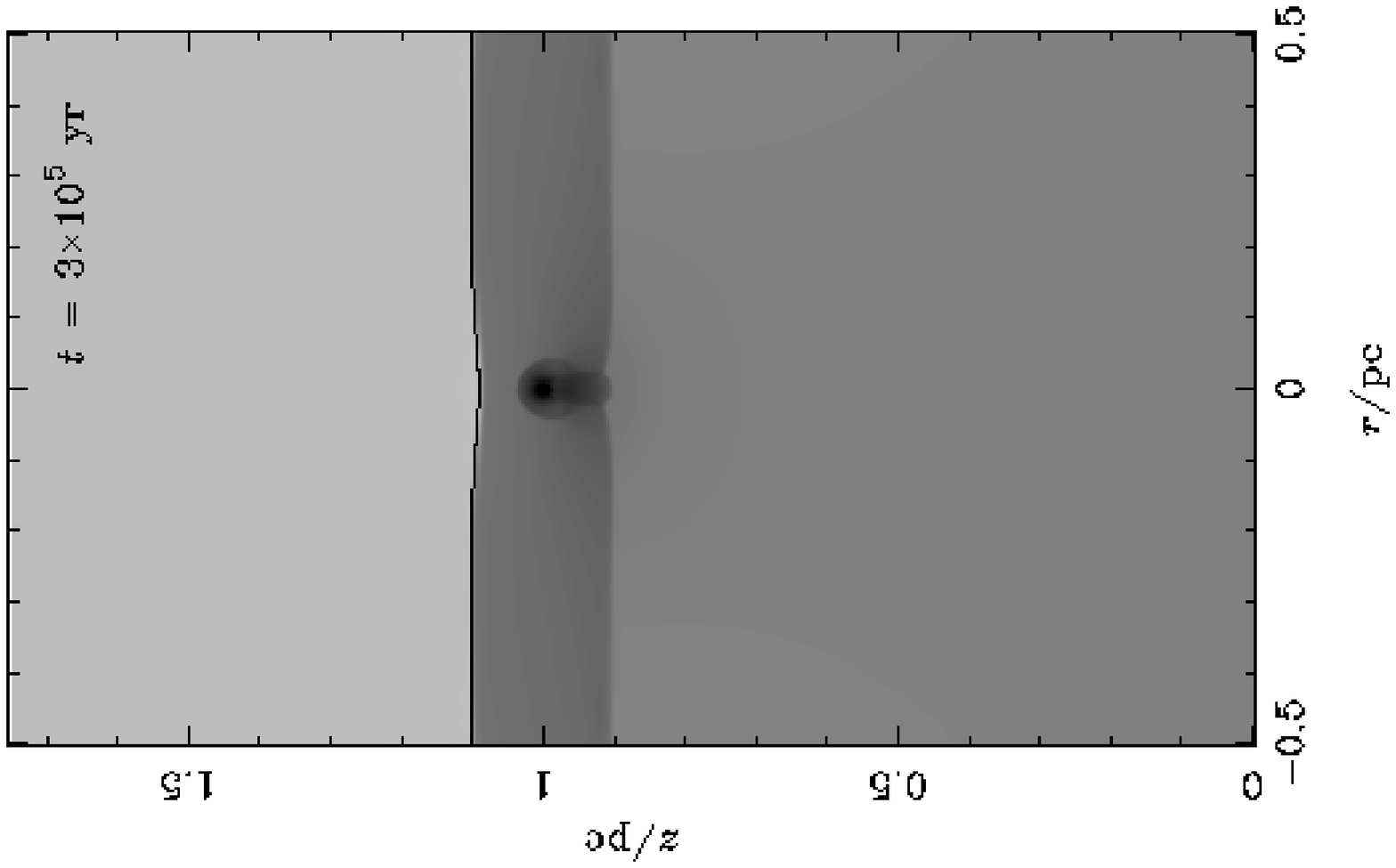}} &
\epsfysize=5cm\rotatebox{270}{\epsffile{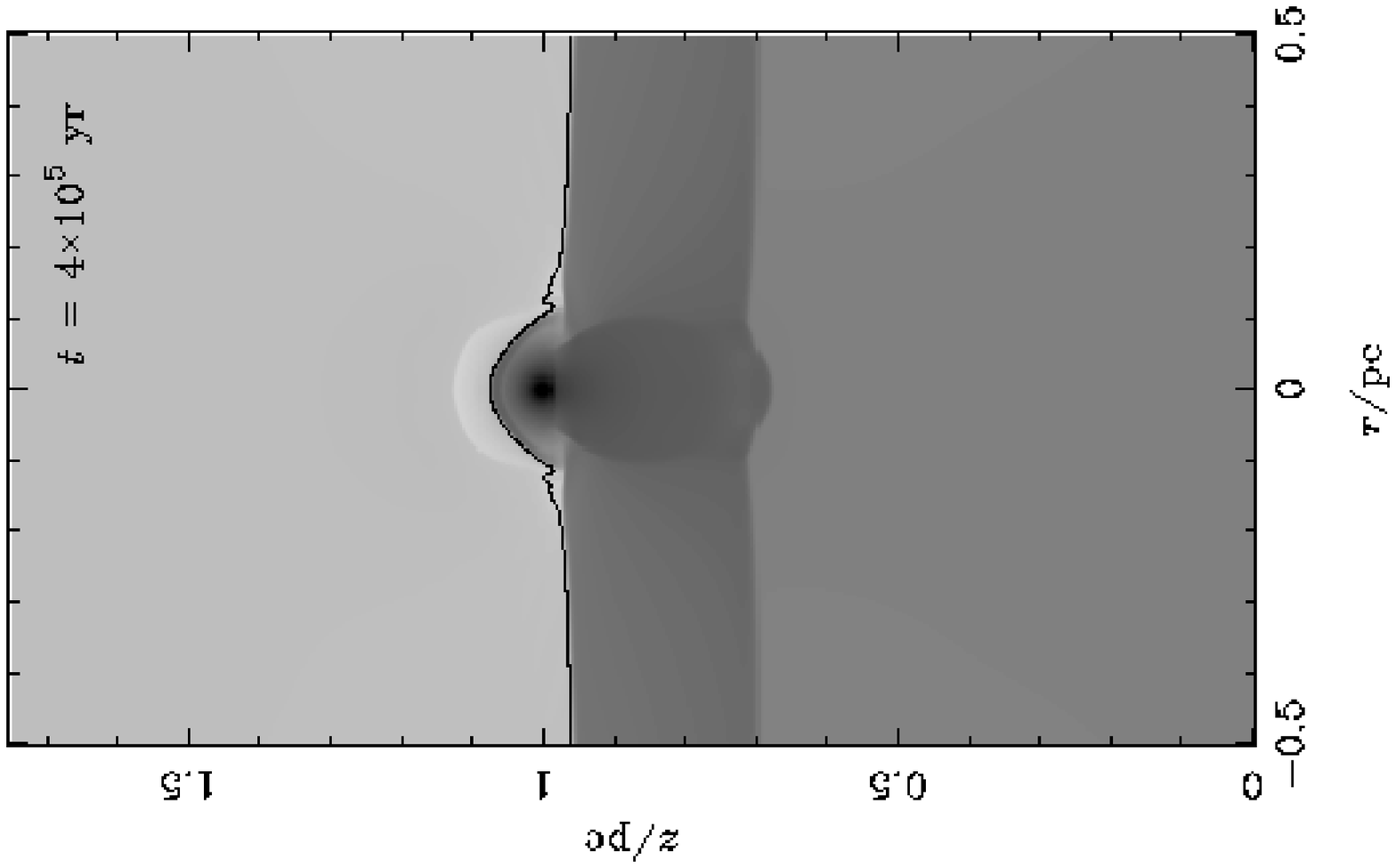}} \\
(d) & (e) & (f) \\
\epsfysize=5cm\rotatebox{270}{\epsffile{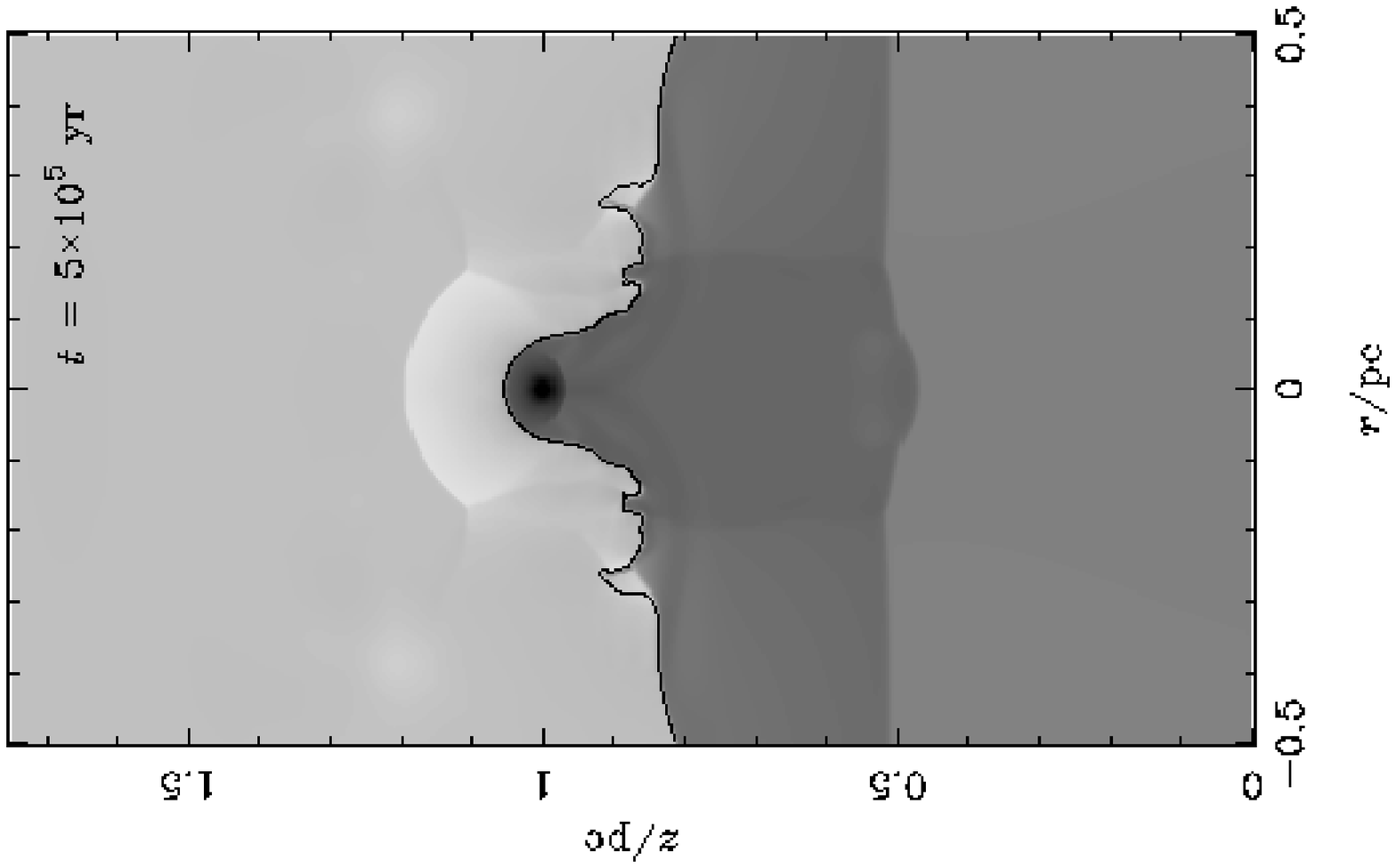}} &
\epsfysize=5cm\rotatebox{270}{\epsffile{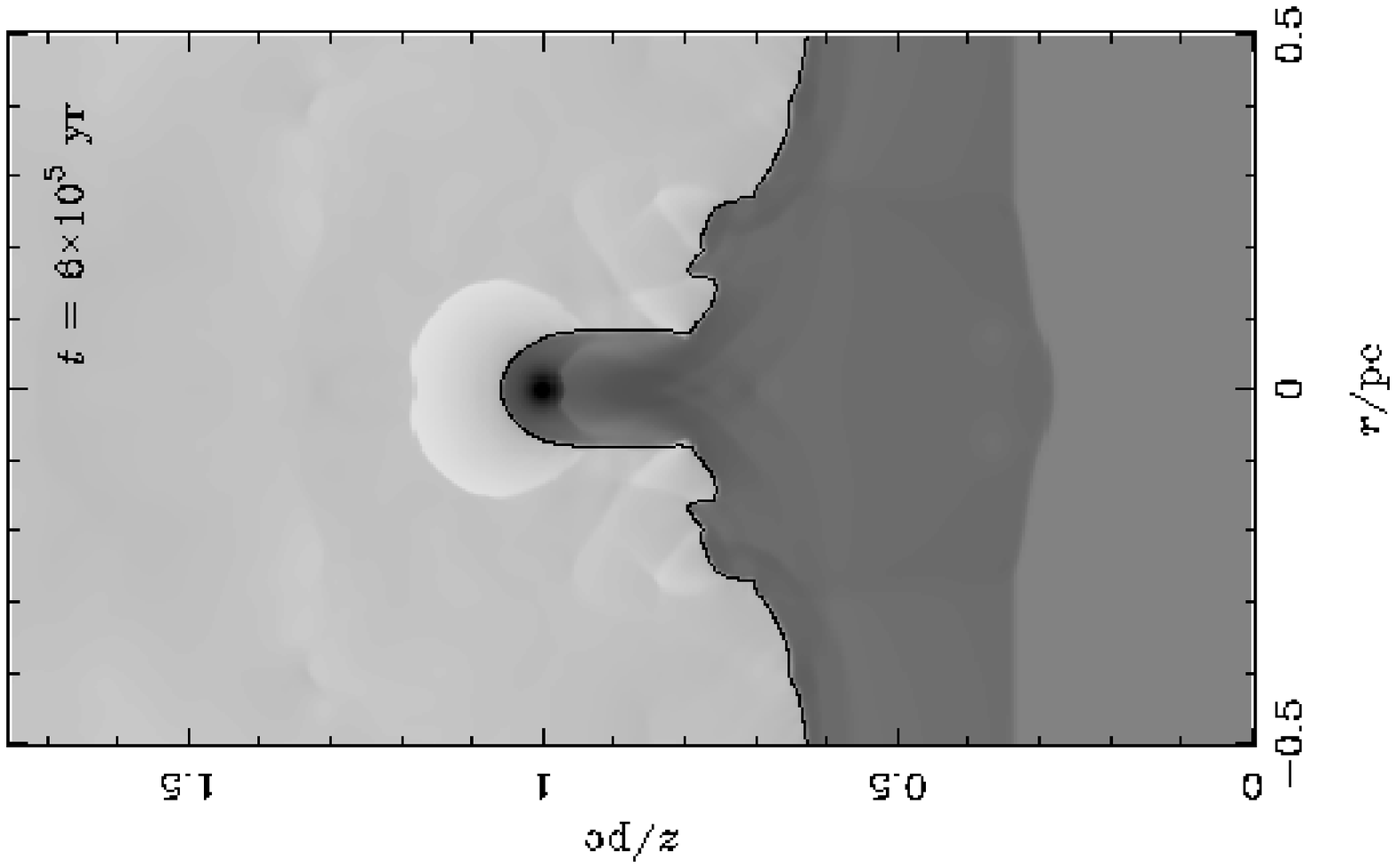}} &
\epsfysize=5cm\rotatebox{270}{\epsffile{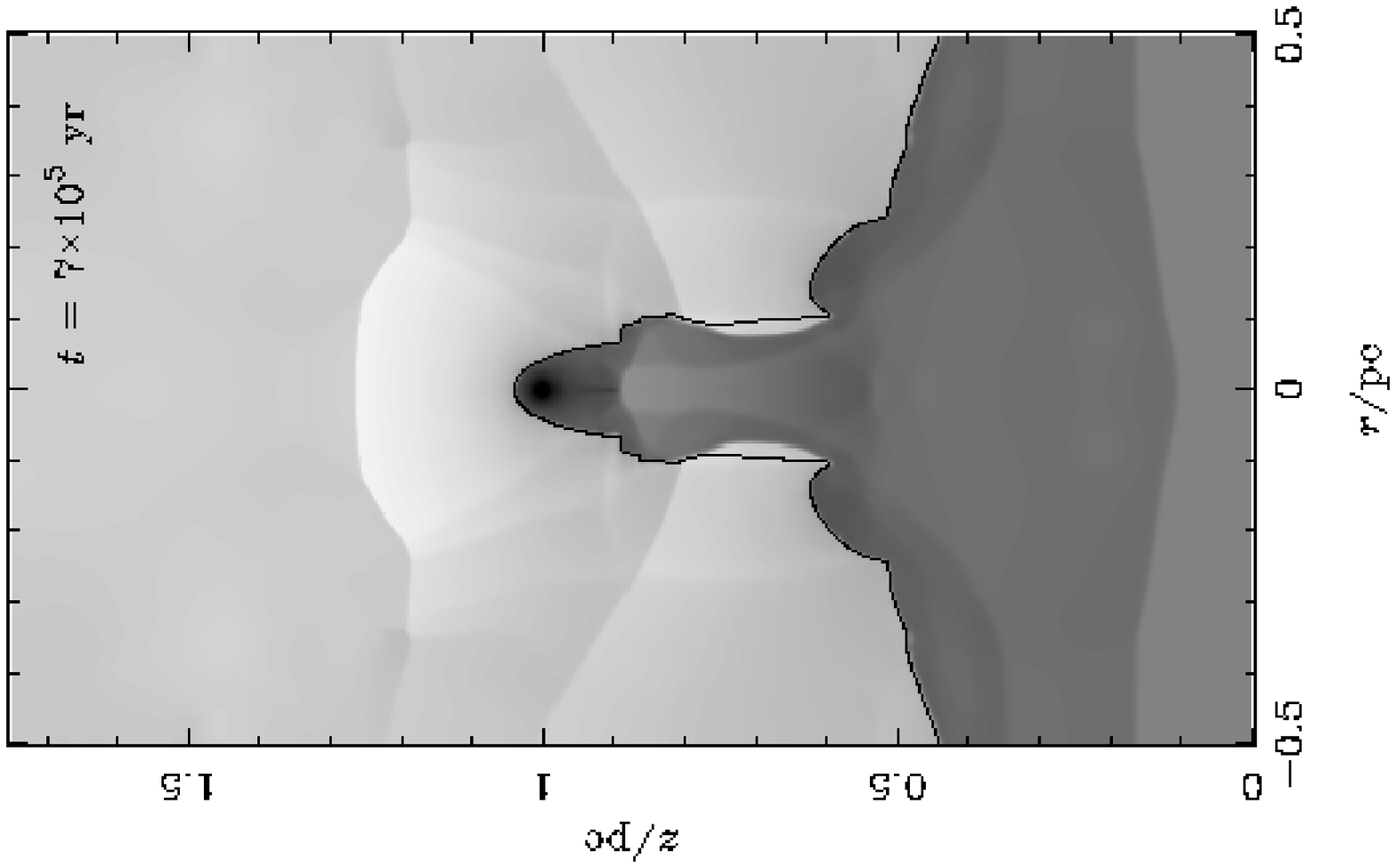}}\\
\multicolumn{3}{c}{\epsfysize=10cm\rotatebox{270}{\epsffile{wedge.ps}}}
\end{tabular}
\end{centering}
\caption{Greyscale plots of log density for Case IV: stationary,
initially uniform gas with a $30\Msun$ gravitational field applied.
The plots are shown at (a) $2\ee5\yr$, (b) $3\ee5\yr$, (c) $4\ee5\yr$,
(d) $5\ee5\yr$, (e) $6\ee5\yr$, and (f) $7\ee5\yr$ after the start of
the simulation.  The density range calibration is shown in the colour
bar.}
\label{f:simgrav}
\end{figure*}

\subparagraph{Case IV:}

In Figure~\ref{f:simgrav}, we take uniform dense gas to nearly fill
the grid, and include a gravitational field equivalent to a
$30$-$\Msun$ core smoothed at a radius of $0.02\parsec$.  In this case
only, the neutral gas is initially at rest.  As the simulation
develops, mass first falls into the gravitational potential, to form a
dense core close to hydrostatic equilibrium.  A shock and ionization
front driven by the impinging ionizing radiation propagate over this
core, generating a column structure.  This column is very long-lived,
tending to form a cometary structure at late times.  The length of
this cometary structure may in reality be limited by the non-plane
parallel ionizing radiation field resulting from diffuse emission and
the multiple ionizing stars \cite{pavea2001}.

This model seems similar to that discussed by White~\etal{} However,
it will be noted that by the time the ionization front reaches the
dense clump, it interacts with the reflection of the main shock front
from the dense core, rather than driving a separate shock itself.
Shocks are always present within the solution, but are a
characteristic of the dynamically disturbed nature of the flow.

Here we have {\it imposed}\/ a separate potential, rather than
calculating the gravity self-consistently, primarily as a
computational convenience.  However, it should be noted that in the
Eagle columns, a reasonable fraction of the gravitational potential
may be generated by protostellar clumps which have effectively
decoupled from the flow.  The small smoothing radius means that the
gravitational field could not be generated by the self-gravity of a
stable isothermal molecular gas core.  It is possible that
protostellar cores may be formed from gravitational instabilities in
the swept up gas shell (Francis \& Whitworth, in preparation), or be
present in the initial molecular cloud.

Note, however, that if the heads of the columns were formed from
individual isothermal cores confined by self-gravity, then these cores
must have been surrounded by significantly hotter, less dense gas if
the overall structure was not to be unstable to gravitational
collapse.  In this case, the source of the $\sim10^4\cm^{-3}$ gas in
the column behind the column is unclear.

\subsection{Discussion}

In each of the cases presented in the present section, structures
strikingly reminiscent of those observed by White
\etal~\shortcite{white99} are maintained within the computational grid
for periods of several $10^5\yr$.  The overall structures observed for
models run for a range of similar parameters were broadly the same,
suggestion that our results are not dependent on the details of our
numerical or physical assumptions.  This suggests that structures seen
in M~16 are in a near-steady state, with the head of the column
confined by the rocket effect at the front and by the ram pressure of
the gas being swept up, by self-gravity, or a combination of the two.
The structures can survive for several times $10^5\yr$, longer than
the shock crossing time of the head of the column, and similar to the
dynamical age of M~16 as a whole.  However, the influence of the
symmetry axis on the stability of these columns means that this
interpretation is tentative, until fully three-dimensional models are
available.

\begin{figure}
\begin{centering}
\epsfysize=8cm\rotatebox{270}{\epsffile{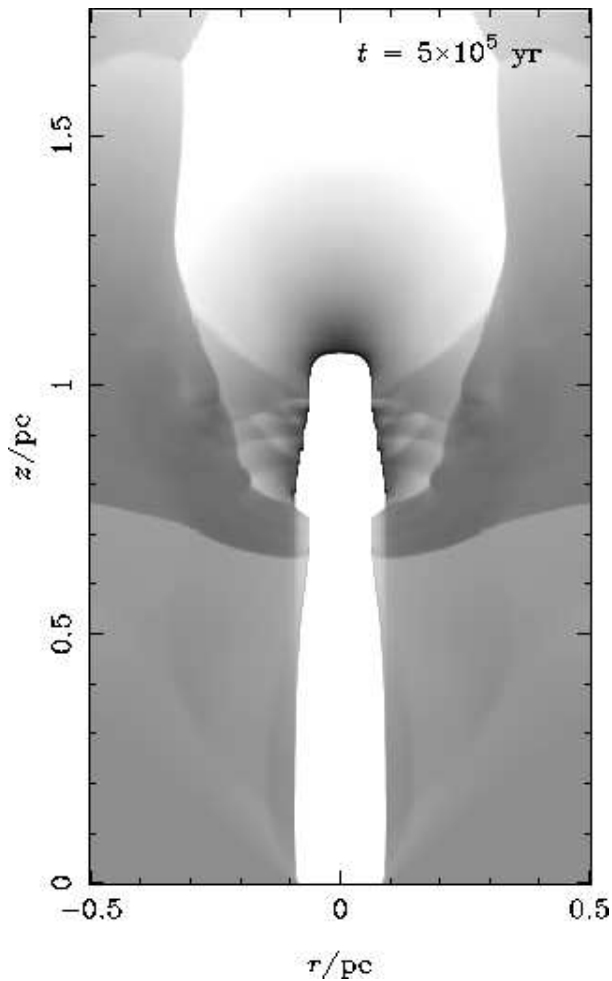}}
\end{centering}
\caption{Greyscale plot of H$\alpha$ emission.  The values are
calculated for the simulation frame shown in
Figure~\protect\ref{f:simwhite}(f), $5\ee5\yr$ after the start of the
simulation.  The greyscale is linear, and saturates at an intermediate
flux.  Note that the emissivity in the symmetry plane is shown here,
rather than the line-of-sight integral.}
\label{f:hab}
\end{figure}

In Figure~\ref{f:hab}, we show the H$\alpha$ emissivity expected from
the gas within Case I after $5\ee5\yr$ from the start of the
simulation, calculated as in Steffen~\etal~\shortcite{stefea97}.  This
image shows the presence of bright rims and strong linear features in
the flow away from the surface.  

Since the structures appear to be close to equilibrium, the
observations contain limited information about the mechanisms by which
they formed: as we have shown, a variety of initial conditions can
lead to very similar final structures.  However, the observations seem
not to {\it require}\/ highly stratified density structures to have
been present in the original molecular cloud.  Indeed, it seems
possible that the columns result from instabilities of the rim of a
champagne-type expanding \HII\ region \cite{tt79,bodea79}.  It is
difficult to make substantial analytic progress in the study of such
surface perturbations, as a result of the nonlocal nature of the
system.  As we have seen in the simulations, density enhancements
resulting from fluctuations in the surface a significant time ago can
result in long term variations in the ionizing flux incident on the
surface a considerable distance from the original perturbation.  While
it is certainly true that the assumption of cylindrical symmetry
enhances the variation of ionizing flux on the axis, it seems likely
that the long lags between perturbation and response inherent in the
system will in general lead to overstabilities of the flows, when
studied on large scales.

The interpretation we have developed, however, has its own
coincidental features.  Most particularly, the mass of shocked
molecular gas at the head of the columns is curiously close to the
Jeans mass.  There is no obvious dynamical reason for the mass of
dense gas to settle on any particular value: indeed, we have seen the
mass of dense gas to be highly variable in our simulations.  A
possible explanation for why the mass is close to the limit of
collapse is that, in fact, the mass at the head increases as the
column erodes, but once it reaches the limit of stability it
collapses.  The dense, self-gravitating globules produced will then no
longer be subject to significant radiation or hydrodynamic forces, and
can be left behind by the further photoevaporation of the column.  In
this manner, an ionization front can lead to stars forming in an
episodic manner over a substantial period, compared to the one-off
radiation-induced collapse of clumps which has been studied previously
\cite[Bok, as referred to by]{oort54,sandea82,bt84,bert89}.  Note that
the ionizing radiation incident on a clump will increase gradually
when the main ionization front is of D-type, rather than suddenly as
often assumed.  The shock which precedes the IF may however result in
a thin gas layer unstable to gravitational
collapse~\cite{el77,whitea94} and may trigger the collapse of any
pre-existing self-gravitating clumps \cite{mw97}.  To develop these
ideas further requires models in which the self-gravity of the
molecular gas is included.

\section{Conclusions}

\label{s:concl}
In this paper, we have presented hydrodynamical models of the
development of photoionized columns, which agree well with the
observations of the columns in M~16.  Our models contrast with the
results of White \etal, who interpret observations of the columns in
M~16 as requiring that they have been exposed to the radiation field
for less than $10^5$ years, the dense material in the head of the
column being a gravitationally-bound condensation from the initial
molecular cloud.

It remains to determine unambiguously how columns such as those in
M~16 form, and how they evolve, if indeed there is any single
mechanism.  It seems likely, however, that these structures are often
long lived components of the environment of massive stars, and of
great importance in the formation of lower mass stars in these
environments, and so they warrant further detailed study.

\section*{Acknowledgments}

The work of RJRW is supported by the PPARC through the award of an
Advanced Fellowship.  We would like to thank the referee, Matt Redman,
for a helpful and constructive report.

\label{lastpage}

\begin{thebibliography}{}

\bibitem[\protect\citename{Andr\'e }1996]{andre96}
Andr\'e P., 1996, Mem.S.A.It., 67, 901

\bibitem[\protect\citename{Axford }1964]{axfo64}
Axford W.I., 1964, ApJ, 140, 112

\bibitem[\protect\citename{Bedijn \& Tenorio-Tagle }1984]{bt84}
Bedijn P.J., Tenorio-Tagle G., 1984, A\&A, 135, 81

\bibitem[\protect\citename{Bertoldi }1989]{bert89}
Bertoldi F., 1989, ApJ, 346, 735

\bibitem[\protect\citename{Bertoldi \& McKee }1990]{bm}
Bertoldi F., McKee C.F., 1990, ApJ, 354, 529

\bibitem[\protect\citename{Bertoldi \etal{} }1993]{bertea93} 
Bertoldi F., McKee C.F., Klein R.I., 1993, in `Massive Stars: Their
Lives in the Interstellar Medium', ASP Conf.\ Ser.\ 35,
Cassinelli J.P., Churchwell E.B., eds., 129

\bibitem[\protect\citename{Bertoldi \& Draine }1996]{bd}
Bertoldi F., Draine B.T., 1996, ApJ, 458, 222

\bibitem[\protect\citename{Bodenheimer \etal{} }1979]{bodea79}
Bodenheimer P., Tenorio-Tagle G., Yorke H.W., 1979, ApJ, 233, 85

\bibitem[\protect\citename{Cant\'o \etal{} }1998]{cantea98}
Cant\'o J., Raga A.C., Steffen W., Shapiro P.R., 1998, ApJ, 502, 695

\bibitem[\protect\citename{Elmegreen \& Lada }1977]{el77}
Elmegreen B.J., Lada C.J., 1977, ApJ, 214, 725

\bibitem[\protect\citename{Falle }1991]{falle}
Falle S.A.E.G., 1991, \mn, 250, 581

\bibitem[\protect\citename{Garc\'\i{}a-Segura \& Franco }1996]{gsf96}
Garc\'\i{}a-Segura G., Franco J., 1996, ApJ, 469, 171

\bibitem[\protect\citename{Giuliani }1979]{giuli79}
Giuliani J.L., 1979, \apj, 233, 280

\bibitem[\protect\citename{Henney \& Arthur }1998]{ha98}
Henney W.J., Arthur S.J., 1998, AJ, 116, 322

\bibitem[\protect\citename{Hester \etal{} }1996]{hester96}
Hester J.J., \etal{}, 1996, AJ, 111, 2349

\bibitem[\protect\citename{Kahn }1958]{kahn58}
Kahn F.D., 1958, Rev.\ Mod.\ Phys., 30, 1058

\bibitem[\protect\citename{Lefloch \& Lazareff }1994]{ll}
Lefloch B., Lazareff B., 1994, A\&A, 289, 559

\bibitem[\protect\citename{Lefloch, Lazareff \& Castets }1997]{llc97}
Lefloch B., Lazareff B., Castets A., 1997, A\&A, 324, 249

\bibitem[\protect\citename{Levenson \etal }2000]{levea00}
Levenson N.A., Graham J.R., McLean I.S., Becklin E.E.,
Figer D.F., Gilbert A.M., Larkin J.E., Teplitz H.I., 
Wilcox M.K., 2000, ApJL, 533, L53

\bibitem[\protect\citename{Megeath \& Wilson }1997]{mw97}
Megeath S.T., Wilson T.L., 1997, AJ, 114, 1106

\bibitem[\protect\citename{Mellema \etal\ }1998]{mellea98}
Mellema G., Raga A.C., Cant\'o J., Lundqvist P., Balick B.,
Steffen W., Noriega-Crespo A., 1998, \ana, 331, 335

\bibitem[\protect\citename{Nelson \& Langer }1997]{nl}
Nelson R.P., Langer W.D., 1997, ApJ, 482, 796

\bibitem[\protect\citename{Oort }1954]{oort54}
Oort J.H., 1954, BAN, 12, 177

\bibitem[\protect\citename{Oort \& Spitzer }1954]{os54}
Oort J.H., Spitzer L., 1954, ApJ, 121, 6

\bibitem[\protect\citename{Pavlakis \etal{} }2001]{pavea2001}
Pavlakis K.G., Williams R.J.R., Dyson J.E., Falle S.A.E.G.,
Hartquist T.W., 2001, A\&A, in press

\bibitem[\protect\citename{Pound }1998]{pound98}
Pound M.W., 1998, ApJL, 493, L113

\bibitem[\protect\citename{Rubin }1968]{rubin68}
Rubin R.H., 1968, ApJ, 153, 761

\bibitem[\protect\citename{Sandford \etal{} }1982]{sandea82}
Sandford M.T., Whitaker R.W., Klein R.I., 1982, ApJ, 260, 183

\bibitem[\protect\citename{Sankrit \& Hester }2000]{sankh00}
Sankrit R., Hester J.J., 2000, ApJ, 535, 847

\bibitem[\protect\citename{Steffen \etal{} }1997]{stefea97}
Steffen W., G\'omez J.L., Williams R.J.R., Raga A.C., Pedlar A., 
1997, MNRAS, 286, 1032

\bibitem[\protect\citename{Sysoev }1997]{sysoev97}
Sysoev N.E., 1997, Astron. Lett., 23, 409

\bibitem[\protect\citename{Tenorio-Tagle }1979]{tt79}
Tenorio-Tagle G., 1979, A\&A, 71, 59

\bibitem[\protect\citename{Vandervoort }1962]{van62}
Vandervoort P.O., 1962, ApJ, 135, 212

\bibitem[\protect\citename{White \etal{} }1999]{white99}
White G.J., \etal, 1999, A\&A, 342, 233

\bibitem[\protect\citename{Whitworth \etal{} }1994]{whitea94}
Whitworth A.P., Bhattal A.S., Chapman S.J., Disney M.J., Turner J.A.,
1994, MNRAS, 268, 291

\bibitem[\protect\citename{Williams }1999]{will99}
Williams R.J.R., 1999, MNRAS, 310, 789

\end{thebibliography}
\end{document}